\documentclass[prd,nofootinbib,english]{revtex4}
\usepackage{amsmath}
\usepackage{amsfonts,color}
\usepackage{amssymb,float,multirow}
\usepackage[utf8]{inputenc}
 \allowdisplaybreaks
\usepackage{color}
\usepackage[unicode=true,pdfusetitle,
 bookmarks=true,bookmarksnumbered=true,bookmarksopen=true,bookmarksopenlevel=3,
 breaklinks=false,pdfborder={0 0 1},backref=false,colorlinks=true]
 {hyperref}
\usepackage{enumitem}

\usepackage{tikz}
\usetikzlibrary{shapes.geometric}
\usetikzlibrary{arrows.meta,arrows}

\begin{document}
\title{String-inspired Teleparallel Cosmology }

\author{Sebastian Bahamonde}
\email{sbahamonde@ut.ee, sebastian.beltran.14@ucl.ac.uk}
\affiliation{Laboratory of Theoretical Physics, Institute of Physics, University of Tartu, W. Ostwaldi 1, 50411 Tartu, Estonia}
\affiliation{Laboratory for Theoretical Cosmology, Tomsk State University of
Control Systems and Radioelectronics, 634050 Tomsk, Russia (TUSUR)}
\affiliation{Department of Mathematics, University College London, Gower Street, London, WC1E 6BT, United Kingdom}

\author{Mihai Marciu}
\email{mihai.marciu@drd.unibuc.ro}
\affiliation{Faculty of Physics, University of Bucharest, 405 Atomistilor, POB MG-11, RO-077125 Bucharest-Magurele, Roumania}
\author{Sergei~D.~Odintsov}
\email{odintsov@ieec.uab.es}
\affiliation{Institut de Ciencies de lEspai (IEEC-CSIC),
    Campus UAB, Carrer de Can Magrans, s/n, 08193 Cerdanyola del Valles, Barcelona, Spain}
    \affiliation{ICREA, Passeig LluAs Companys, 23,
    08010 Barcelona, Spain}
\affiliation{Tomsk State Pedagogical University,
634061 Tomsk, Russia}
\affiliation{Laboratory for Theoretical Cosmology, Tomsk State University of
	Control Systems and Radioelectronics, 634050 Tomsk, Russia (TUSUR)}

\author{Prabir Rudra}
\email{prudra.math@gmail.com, rudra@associates.iucaa.in}
\affiliation{Department of Mathematics, Asutosh College, Kolkata-700026, India.}

\begin{abstract}
The present paper represents an attempt for a very generic string inspired theory of gravitation, based on a stringy action in the teleparallel gravity which includes a specific functional which depends on the scalar field and its kinetic energy, as well as the torsion and boundary terms, embedding also possible effects from the teleparallel Gauss--Bonnet invariants. We focus our study on FLRW cosmology. After we deduce the cosmological equations for the associated generic theory of gravitation, we focus on string inspired couplings which are studied by considering different analytical techniques. The first analytical technique is based on the linear stability theory, by introducing proper dimensionless variables which enables us to study the structure of the phase space and the associated physical effects. In this case, we have obtained different cosmological solutions which correspond to matter and dark energy dominated solutions, achieving a possible transition between matter and dark energy dominated epochs. For each type of cosmological solutions, we have discussed the corresponding physical features, attaining viable constraints for the coupling constants due to dynamical effects. The dynamical study of the physical features included also a numerical analysis by fine--tuning the initial conditions deep into the matter era, obtaining possible trajectories for the effective equation of state for specific coupling functions.
\end{abstract}

\maketitle

\section{Introduction}\label{sec:introduction}
It is rather well-known that the modern theory of the Universe's evolution includes two accelerating epochs: the early-time acceleration and late dark energy epoch. Standard General Relativity (GR) seems to have problems for a natural account of the acceleration eras of the universe evolution. The account of cosmological constant may improve the situation and may give the chance to describe the accelerating universe. However, it brings also a number of problems like fine-tuning, disagreement with some observational data, etc.
Another approach for a natural description of the accelerating universe is the modification of gravity. In fact, different models of modified gravity have been studied recently (for reviews,
 see~\cite{Capozziello:2011et,Nojiri:2017ncd,Nojiri:2010wj}). One of the natural candidates for modified gravity is related to string theory. 
Indeed, it is well-known that higher-order curvature corrections to gravitational action are given by low-energy (super)string effective action~\cite{Fradkin:1985ys,Kawai:1998ab,Antoniadis:1993jc,Tseytlin:1991xk,Brustein:1994kw,Easther:1995ba,Easther:1996yd,Antoniadis:1988vi,Veneziano:1991ek}. The leading order correction in the string low-energy effective action is given by the Gauss-Bonnet invariant which enjoys several remarkable features like being topological invariant in four dimensions. The next-to-leading terms which are of third or fourth-order in curvature invariants depend on the type of string theory under consideration as well as a specific compactification scheme. 

The higher-order curvature corrections also are coupled to scalars (dilaton/modulus) fields~\cite{Gasperini:1994xg,Hwang:2005hb}. These scalar fields appear in the process of compactification from higher to four dimensions. Usually, modulus fields associated with the radii of internal space may be neglected. Then, string effective action contains  GR plus higher-order curvature corrections coupled to dilaton. Such string effective action given in terms of metric tensor and dilaton is often considered as a realistic theory to describe the evolution of the Universe including dark energy~\cite{Elizalde:2007pi,Kanti:1998jd,Maeda:2011zn} and inflation or even early inflationary epoch~\cite{Guo:2009uk,Guo:2010jr,Jiang:2013gza,Kanti:2015pda,Nozari:2017rta,Chakraborty:2018scm,Odintsov:2018zhw,Yi:2018dhl,vandeBruck:2016xvt}.

Recently, a different framework to GR has become popular in the modified gravity community, which is known as Teleparallel gravity (TG), where it is assumed that the manifold contains torsion but the curvature is zero. One of its most popular theories is the so-called $f(T)$ theory, where the action depends on an arbitrary function of the torsion scalar which is a quantity constructed from contractions of the torsion tensor~\cite{Cai:2015emx,Ferraro:2006jd,Ferraro:2008ey}. Geometrically, its description is qualitatively different from that of usual metric gravitation. One of the most important features of $f(T)$ gravity is that for $f(T)=T$, the theory is equivalent to GR. This theory is known as the Teleparallel equivalent of General Relativity (TEGR). In several papers, it has been found that $f(T)$ gravity can describe both the early and late-time eras of cosmology without evoking any cosmological constant~\cite{Bamba:2010wb,Dent:2011zz,Wu:2010av,Hohmann:2017jao}. Moreover, it can alleviate the growing tension of the $H_0$ value~\cite{Nunes:2018xbm}, and also one can obtain bouncing solutions~\cite{Cai:2011tc}. It is also important to mention that the speed of the gravitational waves is equal to the speed of light in this theory~\cite{Cai:2018rzd}. A recent review article on extended cosmography and modified gravity theories, with a wide discussion on $f(T)$ cosmology is given in Ref.~\cite{Capozziello:2019cav}.

Further generalisations of $f(T)$ have been also proposed by different authors in different contexts. One important modification for our paper is the one firstly introduced in~\cite{Kofinas:2014owa}, where the Teleparallel analogue of the modified Gauss-Bonnet theory $f(\mathring{R},\mathring{G})$ was proposed. In this theory, the action depends $f$ on both $T$ and $T_G$ which is similar to the standard Gauss-Bonnet term $\mathring{G}$. In general, this theory is different to $f(\mathring{R},\mathring{G})$, similarly as $f(T)$ gravity is different from $f(\mathring{R})$ gravity~\cite{Bahamonde:2015zma,Bahamonde:2017wwk}. Its cosmology was then studied in~\cite{Kofinas:2014daa,Kofinas:2014aka}, finding that the theory can reproduce different kinds of cosmological eras, giving a consistent unification from early-times inflation to late-times self-acceleration. Later, in~\cite{Bahamonde:2016kba}, the authors showed that by introducing the term $B_G$ which connects $\mathring{G}$ with $T_G$ as $\mathring{G}=-T_G+B_G$, one can further promote the theory to depend on $B_G$ and then be able to formulate a theory containing both $f(T,T_G)$ gravity and $f(\mathring{R},\mathring{G})$ gravity. Since in 4 dimensions $T_G$ and $B_G$ are both topological invariants, one can re-rewrite $T_G$ differently as it was first done in~\cite{Kofinas:2014owa}. This term $T_G$ has four different contractions of the torsion tensor and then, one can decompose $T_G=\sum_i^{4} T_{G_{i}}$. Then, the Gauss-Bonnet term can be written as $\mathring{G}=-\sum_i^{4} T_{G_{i}}+B_G$. It is important to mention that in flat FLRW cosmology, $B_G=0$, and then $\mathring{G}$ coincides with $T_G$. Therefore, in this specific case, the modified corrections coming from the Gauss-Bonnet inclusions are identical. In~\cite{Gonzalez:2019tky}, a theory where all the possible $T_{G_{i}}$ terms were studied. Our formulation relies on unifying the ideas underlying~\cite{Bahamonde:2016kba} and \cite{Gonzalez:2019tky} so that we have a theory depending on a function $f$ with all the 5 Teleparallel Gauss-Bonnet invariants.

Teleparallel gravity gives a rather rich and realistic description of the current universe evolution. This can be also seen in the case where one includes a scalar field in the action. For example, in~\cite{Bahamonde:2015hza,Zubair:2016uhx}, the authors found a Teleparallel theory containing non-minimal couplings between a scalar field and both the scalar torsion $T$ and the boundary term $B$, finding that this theory contains the standard non-minimally coupled theories based on the curvature~\cite{Uzan:1999ch,Bartolo:1999sq}. This was further generalised to quintom models~\cite{Bahamonde:2018miw}, non-local models~\cite{Bahamonde:2017bps,Bahamonde:2017sdo} and specific scalar tensor scenarios~\cite{Bahamonde:2019gjk,Marciu:2017sji,Gecim:2017hmn}, finding again that Teleparallel theories are broader than standard modified theories. Recently, a Teleparallel Horndeski theory was derived which can be written as standard Horndeski plus a correction depending on torsion~\cite{Bahamonde:2019shr}. It is interesting to mention that in~\cite{Bahamonde:2019ipm}, the authors found that, Teleparallel Horndeski can revive the Horndeski terms that were ruled out from gravitational wave observations. It was then concluded that TG has the possibility of constructing broader theories than the standard modified theories by using the corresponding boundary terms connecting both frameworks.

Having in mind the fact that string theory is considered to be the candidate for fundamental theory one can assume that there should exist a Teleparallel formulation of string theory. To check this conjecture at the qualitative level as the first step one can work out with a string effective action to show that it may be re-written equivalently as Teleparallel string effective action. This work is devoted precisely to the development of a Teleparallel string action and Teleparallel string cosmology. Unlike to theories based on the Levi-Civita connection and curvature, in TG, there are five Teleparallel analogues of the Gauss-Bonnet invariant. One expects that for each third and higher-order topological invariant, in curvature-based theories of gravity, there may be even more teleparallel analogs. We show that string effective action up to Gauss-Bonnet term maybe re-written as the equivalent teleparallel string action. However, since one has more possible terms in TG, our action which would corresponds to the first order correction of a string effective action is more general than the standard curvature-based string inspired gravitational theory. Furthermore, for a specific case in our theory, we recover the case studied in~\cite{Elizalde:2007pi}. We also outline how such equivalence may be extended to include higher-order curvature terms. Since that already at the level of second order curvature invariant we get 5 analogs of GB invariant in TG, we can formulate a rather large class of string-inspired teleparallel gravitational theories. This work is devoted to the study the cosmology in such string-inspired teleparallel gravity.

In Sec.~\ref{sec:string} we first briefly introduce Teleparallel gravity and then we propose our theory. After this, we present the modified Friedmann equations in a flat FLRW spacetime. Sec.~\ref{sec:DS} is devoted to studying the cosmology of our model using dynamical system analysis, analyzing the physical properties of our model for the string--inspired couplings. In Sec.~\ref{sec:numerics} we study different models using numerical techniques. We conclude our results in~\ref{sec:Conclusions} and comment about possible routes on how to continue the direction of feasible string theory in the context of Teleparallel gravity. Throughout this paper, we use the geometric units where $c=1$ and the metric signature is $(+---)$, and Latin indices indicate tangents space coordinates whereas Greek indices correspond to spacetime coordinates. Quantities denoted with a overcircle $\circ$ denote that they are computed with the Levi-Civita connection.

\section{String-inspired Teleparallel gravity and FLRW cosmology}\label{sec:string}
Teleparallel gravity is a gravitational theory which assumes a manifold with vanishing curvature but non-zero torsion. The connection in this framework is the so-called Weitzenb\"ock connection. The main ingredient for them is the tetrads $e^a{}_\mu$ whose at each point of the general manifold, gives us bases for vectors on the tangent space. The metric can be reconstructed via $g_{\mu\nu}=\eta_{ab}e^a{}_\mu e^b{}_\nu$. The torsion tensor depends on both the tetrads and the spin connection $w^{a}{}_{b\mu}$ via 
\begin{equation}
T^{a}{}_{\mu\nu} :=  2\Gamma^{a}{}_{[\mu\nu]}\,,
\end{equation}
where $\Gamma^{a}{}_{\mu\nu}$ is the Weitzenb\"ock connection given by
\begin{equation}\label{eq:weitzenbockdef}
\Gamma^{\sigma}{}_{\mu\nu} := e^{a}{}_{\sigma}\partial_\mu e^{a}{}_{\nu} + e^{a}{}_{\sigma}\omega^{a}{}_{b\mu}e^{b}{}_{\nu}\,.
\end{equation}
It is then possible to formulate a theory based on the torsion tensor by considering the following contraction:
\begin{equation}
    T=S^{abc}T_{abc}=\Big[\frac{1}{4}(T^{abc}-T^{bac}-T^{cab})+\frac{1}{2}(\eta^{ac}T^b-\eta^{ab}T^c)\Big]T_{abc}\,,
\end{equation}
which is known as the torsion scalar, that is invariant under both diffemorphisms and under local Lorentz transformations~\cite{Krssak:2015oua}, which has the following action
\begin{equation}
\mathcal{S}_{\rm TEGR}=\frac{1}{2\kappa^2}\int d^4x e\, T+S_{\rm m}\,,\label{action0}
\end{equation}
where $e=\sqrt{-g}=\textrm{det}(e^a{}_\mu)$. Since the curvature is zero, it can be shown that the torsion scalar $T$ and the Ricci scalar $\mathring{R}$ computed with the Levi-Civita connection, are connected via
\begin{equation}\label{TEGR_L}
R=\mathring{R} +T-\frac{2}{e}\partial_{\mu}\left(eT^{\sigma}{}_{\sigma\mu}\right)=0 \quad \Rightarrow \quad \mathring{R} = -T + \frac{2}{e}\partial_{\mu}\left(eT^{\sigma}{}_{\sigma\mu}\right) := -T + B\,.
\end{equation}
From the above equation, one notices that these quantities differ via a boundary term $B$, hence, the action~\ref{action0} gives rise to the same equations as the Einstein-Hilbert action, i.e., the Einstein's field equations. This theory is denoted as the Teleparallel equivalent of General Relativity (TEGR). However, when one modifies the above action, one would get different theories that theories from curvature-based theories. A simple example is to consider the case where we replace $T$ in the action~\ref{action0} for an arbitrary function $f(T)$~\cite{Ferraro:2006jd}. This theory is different from modifying the Einstein-Hilbert action to an arbitrary function, which is the so-called $f(\mathring{R})$ gravity. For the reader interested, see the comprehensive review about TEGR and some of its modifications~\cite{Cai:2015emx,Krssak:2018ywd}. It is important to remark that when one considers modifications of TEGR, one needs to either work with a non-zero spin connection~\cite{Krssak:2015oua} or in a gauge where the spin connection is zero but work in a pure tetrad formalism where the tetrads chose must be consistent with a zero spin connection, which are known as good tetrads~\cite{Tamanini:2012hg}. 

The theory that we are proposing is given by the following action
\begin{equation}
\mathcal{S}=\frac{1}{2\kappa^2}\int d^4x e\, f(\phi,X,T,B,T_{G_1},T_{G_2},T_{G_3},T_{G_4},B_{G})+S_{\rm m}\,,\label{action}
\end{equation}
where $\phi$ is a scalar field, $X= -(1/2)\epsilon \partial_\mu \phi \partial^\mu \phi$ is the kinetic term with $\epsilon$ being a constant that corresponds to either canonical ($\epsilon=1$) or phantom scalar field ($\epsilon=-1$), $T$ is the scalar torsion, $B$ the boundary term, and the other scalars are related to the Teleparallel Gauss-Bonnet invariants that are defined as:
\begin{align}
T_{G_1}&= \delta^{abcd}_{ijkl}K_{a}{}^{i}{}_{e}K_{b}{}^{ej}K_{c}{}^{k}{}_{f}K_{d}{}^{fl} \,,\\
T_{G_2}&=-
2\delta^{abcd}_{ijkl}K_{a}{}^{ij}K_{b}{}^{k}{}_{e}K_{c}{}^{e}{}_{f}K_{d}{}^{fl}\,,\\
T_{G_3}&= 2\delta^{abcd}_{ijkl}K_{a}{}^{ij}K_{b}{}^{k}{}_{e}K_{f}{}^{el}K_{d}{}^{f}{}_{c}\,,\\
T_{G_4}&=2\delta^{abcd}_{ijkl}K_{a}{}^{ij}K_{b}{}^{k}{}_{e}\partial_d K_{c}{}^{el}\,,\\
B_{G} &= \frac{1}{e}\delta^{abcd}_{ijkl} \partial_{a}
\Big[K_{b}{}^{ij}\Big(K_{c}{}^{kl}{}_{,d}+K_{d}{}^{m}{}_{c}K_{m}{}^{kl}\Big)\Big]\,.
\end{align}
where $K_{a}{}^{i}{}_{e}$ is the contortion tensor. The standard Gauss-Bonnet term is related to the Teleparallel ones as:
\begin{eqnarray}
\mathring{G}=\sum_{i=1}^{4}T_{G_i}+B_G=T_{G}+B_G\,.
\end{eqnarray}
The above action contains several theories of gravity in both the Teleparallel framework and curvature-based modified gravitational theories, such as $f(\mathring{R},\phi,X)$~\cite{Beltran:2015hja} by setting $f=f(-T+B,\phi,X)$, or modified Gauss-Bonnet $f(\mathring{R},\mathring{G})$~\cite{Nojiri:2005vv} by choosing $f=f(-T+B,T_{G_1}+T_{G_2}+T_{G_3}+T_{G_4}+B_G)$. Obviously, this theory is also a generalisation of the Teleparallel Gauss-Bonnet theory with $B_G$ studied in~\cite{Bahamonde:2018ibz,Bahamonde:2016kba}.

One can find the field equations for this theory by taking variations with respect to the tetrad field. Since our action depends on many variables, the final expression of these equations is quite long and cumbersome, so that, we will not write them here explicitly. The easiest way to find them out is to use the expressions computed in~\cite{Bahamonde:2016kba}, where the field equations of $f(T,B,T_{G},B_G)$ were found. In our case, the action depends on all the possible Teleparallel Gauss-Bonnet invariants, so that, the field equations would be slightly different from the ones reported in~\cite{Bahamonde:2016kba}. However, it is not difficult to obtain them since in the appendix of~\cite{Bahamonde:2016kba}, the authors also found the variations of each quantity $\delta T_{G_{i}}$ and $\delta B_{G}$ in a separate way. Since we are interested in studying cosmology, it is easier to work out using the minisuperspace of flat FLRW cosmology constructed from the point-like Lagrangian instead of working directly with the field equations. It is well-known that it is equivalent to compute the equations from the field equations and then assume FLRW cosmology, or to compute the FLRW equations from the point-like Lagrangian. Moreover, this procedure commutes for any spherically symmetric space-time. When one considers other spacetimes with fewer symmetries, this procedure is no longer equivalent. Therefore, we will not write down the field equations of our theory here.

The scalar field equation is much easier and simpler to write, which can be found by taking variations with respect to the scalar field, yielding the following modified Klein-Gordon equation,
\begin{eqnarray}
\frac{1}{e}\epsilon\,\partial_{\mu}\Big(ef_{X}g^{\mu\nu}\partial_{\nu}\phi\Big)+f_{\phi}=0\,,\label{fieldeq2}
\end{eqnarray}
where subscripts denote derivatives. It is easy to check that the field equation satisfies the standard conservation equation $\mathring{\nabla}_\mu H^{\mu\nu}=\kappa^2 \mathring{\nabla}_\mu \Theta^{\mu\nu}=0$, since the matter is minimally coupled to the gravitational sector.
Let us consider flat FLRW cosmology in Cartesian coordinates given by the line element $ds^2=dt^2-a(t)^2(dx^2+dy^2+dz^2)$. The diagonal tetrad $e^{a}{}_\mu=\textrm{diag}(1,a(t),a(t),a(t))$ reproduces the FLRW metric and it is a good tetrad in the sense that the antisymmetric part of the field equations is zero. For this space-time, we have 
\begin{eqnarray}
T=-6H^2\,, \quad B=-6(3H^2+\dot{H})\,,\quad T_{G_3}=24H^4\,,\quad T_{G_4}=24H^2\dot{H}\,,\quad B_G=T_{G_1}=T_{G_2}=0\,.
\end{eqnarray}
Here, $H=\dot{a}/a$ is the Hubble parameter and dots denote differentiation with respect to time. Clearly, $\mathring{R}=-T+B=-6(2H^2+\dot{H})$ and $\mathring{G}=T_{G_1}+T_{G_2}+T_{G_3}+T_{G_4}+B_G=24H^2(H^2+\dot{H})$ are recovered as expected.
Since $B_G=T_{G_1}=T_{G_2}=0$, the important terms in the action \eqref{action} are  $f(\phi,X,T,B,T_{G_3},T_{G_4})$. Then, for the general model, the modified FLRW equations read
\begin{eqnarray}
12H^2\left(3 H^2-  \dot{H}\right) f_{T_{G_4}}-48 H^4 f_{T_{G_3}}+6 H^2 f_{T}-\frac{1}{2} \epsilon  \dot{\phi}^2 f_{X}+\frac{1}{2} f+12 H^3 \dot{f}_{{T}_{G_4}}\nonumber\\
-3 H \dot{f}_{B}+3 f_B\left(\dot{H}+3 H^2\right)=\kappa ^2 \rho \,,\label{FRW1}\\
2 H \left(24 H (\dot{H}+H^2) f_{T_{G_3}}-\dot{f}_{T}-2 H \ddot{f}_{T_{G_4}}\right)-36 H^2 \left(\dot{H}+H^2\right) f_{T_{G_4}}-2 \left(\dot{H}+3 H^2\right) f_{T}-\frac{1}{2} f\nonumber\\
-\ddot{f}_B+3f_B\left(\dot{H}+3 H^2\right)+16 H^3 \dot{f}_{T_{G_3}}-8 H \dot{f}_{T_{G_4}} \left(\dot{H}+3 H^2\right)=-\kappa ^2 p\,,\label{FRW2}
\end{eqnarray}
where dots denote differentiation with respect to the cosmic time and again, subscripts denote differentiation. It is easy to see that the theory studied in~\cite{Gonzalez:2019tky} is contained in the above equations. The modified Klein Gordon equation \eqref{fieldeq2} becomes
\begin{equation}
 \ddot{\phi} f_{X}+3  H \dot{\phi} f_{X}+\frac{1}{\epsilon}f_{\phi}+ \dot{f}_X \dot{\phi}=0\,.\label{KG1}
\end{equation}
These cosmological equations are quite general. For example, we can construct the standard modified Gauss-Bonnet theory by taking $f=f(-T+B,T_{G_3}+T_{G_4})=f(\mathring{R},\mathring{G})$~\cite{Cognola:2006eg}.  Motivated by non-minimal theories between the scalar field and the gravitational sector, we will concentrate our study in the following form of the function $f$:
\begin{eqnarray}\label{model1}
f(\phi,X,T,B,T_{G_3},T_{G_4})=T+\kappa^2\Big[2X-2V(\phi)+F_1(\phi)T+F_2(\phi)B+F_3(\phi)T_{G_3}+F_4(\phi)T_{G_4}\Big]\,.\label{f}
\end{eqnarray}
The theory $F_3=F_4=0$ was studied in \cite{Bahamonde:2015hza} with $F_1=\xi \phi^2$ and $F_2=\chi \phi^2$ and then generalised for any $F_1$ and $F_2$ in \cite{Zubair:2016uhx}. The latter theory is also connected to the Teleparallel dark energy~\cite{Geng:2011aj,Xu:2012jf}. Furthermore, the scalar Gauss-Bonnet theory with a coupling $f(\phi)\mathring{G}$ studied in~\cite{Nojiri:2005vv} can be recovered by setting $F_1=F_2=0$ and $F_3=F_4=f(\phi)$.  Note that we have a $+X=-(1/2)\epsilon\,(\partial_\mu \phi)^2$ in the Lagrangian and also a term like $+T$ to have a canonical scalar field when $\epsilon=1$ (due to the signature of the metric). One can also modify this Lagrangian with a $-T$ and with a $-X$ to also have a canonical scalar field. 

By replacing \eqref{f} into the modified FLRW equations \eqref{FRW1}-\eqref{FRW2} we obtain
\begin{eqnarray}
3 H^2 \left(1+\kappa^2F_1\right)&=& \kappa ^2 \Big[\rho +V(\phi)+\frac{1}{2} \epsilon \, \dot{\phi}^2+3  H \dot{\phi}F'_2+36 H^4 (F_3-F_4)-12  H^3 \dot{\phi}F'_4\Big]\,,\label{FFRW1}\\
(3 H^2+2\dot{H})\left(1+\kappa^2F_1\right)&=&\kappa^2\Big[-p+V(\phi)-\frac{1}{2}\epsilon\,\dot{\phi}^2+\dot{\phi}^2 (F''_2-4 H^2 F''_4)+\dot{\phi}\left(8 H^3 \left(2 F'_3-3 F'_4\right)-2 H F'_1\right)\nonumber\\
&&+\ddot{\phi}\left(F'_2-4 H^2 F'_4\right)+12 H^2(3H^2+4 \dot{H}) (F_3-F_4)-8 H \dot{H} \dot{\phi}F'_4\Big]\,,\label{FFRW2}
\end{eqnarray}
where primes denotes differentiation with respect to the scalar field. For this model, the modified Klein-Gordon equation \eqref{KG1} yields
\begin{equation}
    3 H^2 \left(F'_1+3 F'_2-4 \dot{H} F'_4\right)+3 \dot{H} F'_2-12 H^4 F'_3+3 \epsilon  H \dot{\phi}+V'(\phi)+\epsilon  \ddot{\phi}=0\,,\label{KKG1}
\end{equation}

We can rewrite the modified FLRW equations as follows
\begin{eqnarray}
    3H^2&=&\kappa^2(\rho+\rho_{\rm modified})\,,\\
      3H^2+2\dot{H}&=&-\kappa^2(p+p_{\rm modified})\,,
\end{eqnarray}
where we have defined the energy density and pressure for the modifications of General Relativity as
\begin{eqnarray}
\rho_{\rm modified}&=&-3H^2F_1(\phi)+V(\phi)+\frac{1}{2} \epsilon \, \dot{\phi}^2+3  H \dot{\phi}F'_2+36 H^4 (F_3-F_4)-12  H^3 \dot{\phi}F'_4\,,\label{rho}\\
p_{\rm modified}&=&-\Big[-(3H^2+2\dot{H})F_1(\phi)+V(\phi)-\frac{1}{2}\epsilon\,\dot{\phi}^2+\dot{\phi}^2 (F''_2-4 H^2 F''_4)+\dot{\phi}\left(8 H^3 \left(2 F'_3-3 F'_4\right)-2 H F'_1\right)\nonumber\\
&&+\ddot{\phi}\left(F'_2-4 H^2 F'_4\right)+12 H^2(3H^2+4 \dot{H}) (F_3-F_4)-8 H \dot{H} \dot{\phi}F'_4\Big]\,.\label{p}
\end{eqnarray}
Then, one can define the effective equation of state parameter as
\begin{equation}
    w_{\rm eff}=\frac{p_{\rm total}}{\rho_{\rm total}}= \frac{\omega\rho+p_{\rm modified}}{\rho+\rho_{\rm modified}}\,,\label{weff}
\end{equation}
where we have also assumed a barotropic equation of state for the matter $p=\omega\rho$.
The continuity equation for the matter fluid is given by,
\begin{equation}\label{continuity}
\dot{\rho}+3\frac{\dot{a}}{a}\left(\rho+p\right)=0\,.
\end{equation}
Inspired by string theory we will also assume that all the functions are given by~\cite{Fradkin:1985ys,Gross:1986mw,Metsaev:1987ju,Mavromatos:2000az}
\begin{eqnarray}
F_i(\phi)=c_i \alpha e^{\phi/\phi_0}\,,\label{F1}
\end{eqnarray}
where $\alpha$ is the string expansion and $c_i$ and $\phi_0$ are constants. For the specific case where $c_1=c_2=0$ and $c_3=c_4=c$, we recover the leading string expansion term $\mathcal{L}_c= c \alpha e^{\phi/\phi_0}\mathring{G}$ that was studied in \cite{Elizalde:2007pi}.

\section{Dynamical system analysis for string-inspired Teleparallel cosmology}\label{sec:DS}

In this section we will analyse the model described by \eqref{FRW1}-\eqref{FRW2} with the coupling functions being exponential-type. For a comprehensive review about dynamical systems in cosmology, see~\cite{Bahamonde:2017ize}. Similarly as in \cite{Bahamonde:2015hza}, let us introduce the following dimensionless variables
\begin{eqnarray}
\sigma ^2=\frac{\kappa ^2 \rho }{3 H^2}\,,\quad x^2=\frac{\kappa ^2 \dot{\phi}^2}{6 H^2}\,,\quad y^2=\frac{\kappa ^2 V(\phi)}{3 H^2}\,,\quad
z= e^{\frac{\phi }{3\phi_0}}\,,\quad v=4\sqrt{2}H z\,.\label{dimensio}
\end{eqnarray}
By using these dimensionless variables and setting $\kappa=1$ for simplicity, the first Friedmann equation \eqref{FRW1} becomes
\begin{equation}
1=\alpha z^3 \Big(\sqrt{6}  c_2 x \phi_0^{-1}-  c_1\Big)+\frac{\alpha  v^2 z}{8} \left(3(c_3-c_4)-\sqrt{6}\phi_0^{-1} c_4 x\right)+\sigma^2+\epsilon\,  x^2+y^2\,,\label{constrs}
\end{equation}
which reduces the dimensionality of the dynamical system from five to four. For simplicity we will assume  an exponential potential such that
\begin{equation}\label{potentialchoice}
    V(\phi)=V_0 e^{-\lambda \phi}\,,
\end{equation}
where $\lambda$ is a constant. The dimensionless variables~\eqref{dimensio} can be related to the modified scalar field Gauss-Bonnet case studied in~\cite{Nojiri:2005vv}. To recover this case, the coupling functions must be $F_1=F_2=0$ and $F_3=F_4=2f(\phi)$ and the parameters $\lambda=2/\tilde{\phi}_0$ and $\phi_0=\tilde{\phi}_0/2$, where $\tilde{\phi}_0$ is the quantity introduced in~\cite{Nojiri:2005vv}. Furthermore, the dynamical system studied in that paper introduced the dimensionless variables
\begin{equation}
    X=\frac{\dot{\phi}}{H}\,,\quad Z=H^2f'(\phi)\,.
\end{equation}
Our dimensionless variables are related to the above by the following relationships,
\begin{align}
    X=\sqrt{6}x\,, \quad Z= \frac{f_0 }{16 \tilde{\phi}_0}v^2 z\,,
\end{align}
whereas, in the latter paper $\rho=0$, so that $\sigma=0$. It should be noted that for this special case, in our dimensionless variables, it is easier to introduce another variable $V=v^2 z$, since only that combination appears. For the general case, the dimensionless variables chosen in~\cite{Nojiri:2005vv} are not the best option to choose since as it can be seen from~\eqref{constrs}, when $c_1,c_2\neq0$ ($F_1,F_2\neq0$), there is an extra term where only z appears which is not multiplied by $v^2$.

Now, let us study the general case. By introducing $N=\log(a)$, we can write the dynamical system of the model as follows
\begin{eqnarray}
\frac{dx}{dN}&=&\frac{1}{A}\Big[64 \alpha ^2 z^6 \left(9 \sqrt{2} c_2\phi_0 (c_1+c_2) x-\sqrt{3} c_1\phi_0^2 (2 c_1+3 c_2)-6 \sqrt{3} c_2^2 x^2\right)+y^2 \Big(64 \alpha \phi_0^2 z^3 \left(\sqrt{3} (2 c_1 \lambda \phi_0-3 c_2)-3 \sqrt{2} c_2 \lambda  x\right)\nonumber\\
&&+24 \alpha \phi_0^2 v^2 z \left(\sqrt{3} (-4 c_3 \lambda \phi_0+4 c_4 \lambda \phi_0+c_4)+3 \sqrt{2} c_4 \lambda  x\right)+64\phi_0^3 \left(2 \sqrt{3} \lambda -3 \sqrt{2} \epsilon  x\right)\Big)\nonumber\\
&&+8 \alpha ^2 v^2 z^4 \left(-3 \sqrt{2}\phi_0 x (5 c_1 c_4+5 c_2 c_3+3 c_2 c_4)+\sqrt{3}\phi_0^2 (c_1 (14 c_3-15 c_4)+27 c_2 (c_3-c_4))+12 \sqrt{3} c_2 c_4 x^2\right)\nonumber\\
&&+64 \alpha \phi_0 z^3 \left(\sqrt{3} \epsilon \phi_0 (4 c_1+9 c_2) x^2+-\sqrt{3}\phi_0 (2 c_1+3 c_2)-3 \sqrt{2} c_1 \epsilon \phi_0^2 x-6 \sqrt{2} c_2 \epsilon  x^3\right)+192 \sqrt{2} \epsilon \phi_0^3 x \left(\epsilon  x^2-1\right)\nonumber\\
&&+3 \alpha ^2 v^4 z^2 \left(\sqrt{3}\phi_0^2 \left(-4 c_3^2+7 c_3 c_4-3 c_4^2\right)+\sqrt{2} c_4\phi_0 (7 c_3-6 c_4) x-2 \sqrt{3} c_4^2 x^2\right)\nonumber\\
&&+8 \alpha \phi_0 v^2 z \left(27 \sqrt{2} \epsilon \phi_0^2 (c_3-c_4) x-\sqrt{3} \epsilon \phi_0 (8 c_3+9 c_4) x^2+\sqrt{3}\phi_0 (2 c_3-3 c_4)+6 \sqrt{2} c_4 \epsilon  x^3\right)\Big]\,,\label{dxN}\\
\frac{dy}{dN}&=&\frac{y}{A}\Big[-8 \sqrt{6} \alpha  \epsilon \phi_0 x z \left((4 c_3-3 c_4) v^2-8 (2 c_1+3 c_2) z^2\right)+3 \Big(-8 \alpha  v^2 z \left(\alpha  z^3 (c_1 c_4+c_2 (c_3+3 c_4))+3 \epsilon \phi_0^2 (c_3-c_4)\right)\nonumber\\
&&+64 \left(\alpha ^2 c_2 (c_1+3 c_2) z^6+\alpha  c_1 \epsilon \phi_0^2 z^3+\epsilon \phi_0^2\right)-8\phi_0 y^2 \left(8 \alpha  c_2 \lambda  z^3-\alpha  c_4 \lambda  v^2 z+8 \epsilon \phi_0\right)+\alpha ^2 c_3 c_4 v^4 z^2\Big)\nonumber\\
&&-48 \epsilon  x^2 \left(8 \alpha  c_2 z^3-\alpha  c_4 v^2 z-4 \epsilon \phi_0^2\right)\Big]-\sqrt{\frac{3}{2}} \lambda  x y\,,\label{dyN}\\
\frac{dz}{dN}&=&\sqrt{\frac{2}{3}}\phi_0^{-1}x z\,,\label{dzN}\\
\frac{dv}{dN}&=&-\frac{v}{A}\Big[-8 \sqrt{6} \alpha    \epsilon \phi_0 x z \left(v^2 (4 c_3-3 c_4 (\omega +1))-8 z^2 (2 c_1-3 c_2 (\omega -1))\right)+3 \Big(-8 \alpha   ^2 v^2 z \Big(\alpha  z^3 (c_1 c_4+c_2 (c_3+3 c_4))\nonumber\\
&&+3 (\omega +1) \epsilon \phi_0^2 (c_3-c_4)\Big)+64 \left(\alpha ^2 c_2  ^2 (c_1+3 c_2) z^6+\alpha  c_1  ^2 (\omega +1) \epsilon \phi_0^2 z^3+(\omega +1) \epsilon \phi_0^2\right)\nonumber\\
&&-8\phi_0 y^2 \left(8 \alpha  c_2   \lambda  z^3-\alpha  c_4   \lambda  v^2 z+8 (\omega +1) \epsilon \phi_0\right)+\alpha ^2 c_3 c_4  ^2 v^4 z^2\Big)-48 \epsilon  x^2 \left(8 \alpha  c_2 z^3-\alpha  c_4 v^2 z+4 (\omega -1) \epsilon \phi_0^2\right)\Big]\nonumber\\
&&+\sqrt{\frac{2}{3}}\phi_0^{-1}v x\,,\label{dvN}
\end{eqnarray}
where for simplicity we have defined the quantity
\begin{equation}
    A=\sqrt{2}\phi_0 \left(64 \left(2 \alpha  c_1 \epsilon \phi_0^2 z^3+3 \alpha ^2 c_2^2 z^6+2 \epsilon \phi_0^2\right)-16 \alpha  v^2 z \left(3 \alpha  c_2 c_4 z^3+6 \epsilon \phi_0^2 (c_3-c_4)-\sqrt{6} c_4 \epsilon \phi_0 x\right)+3 \alpha ^2 c_4^2 v^4 z^2\right)\,.
\end{equation}
The effective state parameter defined in~\eqref{weff} in our dimensionless variables~\eqref{dimensio} becomes
\begin{eqnarray}
w_{\rm eff}&=&\frac{3}{A}\Big[-3 \Big(\alpha  z \left(16 v^2 \left(\alpha  z^3 (c_1 c_4+c_2 c_3)+3 \epsilon  \phi_0^2 (c_4-c_3)\right)-64 \alpha  c_2 (2 c_1+3 c_2) z^5+\alpha  c_4 (3 c_4-2 c_3) v^4 z\right)\nonumber\\
&&+16 \phi_0 y^2 \left(8 \alpha  c_2 \lambda  z^3-\alpha  c_4 \lambda  v^2 z+8 \epsilon  \phi_0\right)\Big)-64 \sqrt{6} \alpha  \epsilon  \phi_0 x z \left(c_3 v^2-2 (2 c_1+3 c_2) z^2\right)\nonumber\\
&&+96 \epsilon  x^2 \left(-8 \alpha  c_2 z^3+\alpha  c_4 v^2 z+4 \epsilon  \phi_0^2\right)\Big]\,.
\end{eqnarray}

\par 
Furthermore, we can introduce the corresponding deceleration 
\begin{equation}
    q=-1-\frac{\dot{H}}{H^2}\,,
\end{equation}
and the statefinder parameters $\{r,s\}$ \cite{Alam:2003sc,Sahni:2002fz}:
\begin{equation}
    r=\frac{\ddot{H}}{H^3}-3q-2=\frac{d}{dN}\left(\frac{\dot{H}}{H^2}\right)+2\left(\frac{\dot{H}}{H^2}\right)^2+3\left(\frac{\dot{H}}{H^2}\right)+1\,,
\end{equation}
\begin{equation}
    s=\frac{r-1}{3\left(q-\frac{1}{2}\right)}\,.
\end{equation}
The expression for the quantity $\left[\frac{\dot{H}}{H^2}\right]$ can be obtained from the relation for the effective state parameter, considering
\begin{equation}
    w_{\rm eff}=-1-\frac{2}{3}\left(\frac{\dot{H}}{H^2}\right)\,.
\end{equation}
\par 
The dynamical system~\eqref{dxN}-\eqref{dvN} has 26 critical points, but only 12 are physical in the sense that ensures the conditions $v\geq0$ and $y\geq0$ which means that we will assume that $H\geq0$ (expanding universes) and $V\geq0$. Table~\ref{Table1} shows all the physical critical points. We have further assumed that $\alpha\neq0$ and $\lambda\neq0$, for the study.

\begin{table}[H]
\centering
 \centering
\resizebox{15cm}{!}{\begin{tabular}{||c | c | c | c| c||} 
 \hline
 Cr.P. & $x$ & $y$ & $z$ & $v$ \\ [0.5ex] 
 \hline\hline
 $P_1$ & 0 & 0 & 0&0 \\ [0.5ex]
 \hline
 $P_{2,\pm}$ & $\pm\displaystyle\frac{1}{\sqrt{\epsilon}}$ & 0 & 0&0 \\ [0.5ex]
 \hline
 $P_{3,\pm}$ & $\displaystyle \sqrt{\frac{3}{2}}\lambda^{-1}$ & $\pm \displaystyle \sqrt{\frac{3\epsilon}{2}}\lambda^{-1}$ & 0&0 \\ [0.5ex]
 \hline
 $P_{4}$ & $\displaystyle\frac{\lambda }{\sqrt{6} \epsilon }$ & $\sqrt{1-\frac{\lambda ^2}{6 \epsilon }}$ & 0&0 \\ [0.5ex]
 \hline
  $P_{5}$ & $0$ & $0$ & $-\displaystyle\sqrt[3]{\displaystyle\frac{1}{\alpha c_1}}$&0 \\ [0.5ex]
 \hline
  $P_{6}$ & $0$ & $\sqrt{\frac{\alpha  z^3 (2 c_1 c_3-3 c_1 c_4+9 c_2 c_3-9 c_2 c_4)-c_3}{c_3 (3 \lambda  \phi_0-1)-3 c_4 \lambda  \phi_0}}$ & Any &$2 \sqrt{-\frac{2 \lambda  \phi_0-2 \alpha  z^3 (-c_1\lambda \phi_0+c_1+3 c_2)}{\alpha  z (-3 c_3 \lambda  \phi_0+c_3+3 c_4 \lambda  \phi_0)}}$ \\ [0.5ex]
 \hline
  $P_{7,\pm}$ & $0$ & $0$ & $\pm \displaystyle\sqrt[3]{\pm\frac{c_3}{\alpha  (2 c_1 c_3-3 c_1 c_4+9 c_2 c_3-9 c_2 c_4)}}$ &$2\sqrt{\frac{2(c_1+3c_2)}{c_3}}\sqrt[3]{\pm\frac{c_3 }{(\alpha  (2 c_1 c_3-3 c_1 c_4+9 c_2 c_3-9 c_2 c_4))}}$  \\ [0.5ex]
  \hline
 $P_{8,\pm}$ & $0$ & $\sqrt{\frac{c_1+3 c_2}{-c_1\lambda \phi_0+c_1+3 c_2}}$ & $\pm\sqrt[3]{\pm\frac{\lambda  \phi_0}{\alpha  (-c_1 \lambda \phi_0+c_1+3 c_2)}}$  & $0$ \\ [0.5ex]
 \hline
  \end{tabular}}
 \caption{Critical points for the exponential coupling model.} \label{Table1}
 \end{table}

\begin{table}[H]
\centering
 \centering
\begin{tabular}{||c| c||} 
 \hline
 Cr.P. & Eigenvalues \\ [0.5ex] \hline\hline
 $P_1$ & $[-\frac{3}{2},-\frac{3}{2},\frac{3}{2} ,0]$ \\ [0.7ex]\hline
  $P_{2\pm}$ & $\big[\pm \frac{\sqrt{2}}{\sqrt{3 \epsilon } \phi _0},3,\frac{6 \sqrt{\epsilon } \mp \sqrt{6} \lambda }{2 \sqrt{\epsilon }},\frac{\pm \sqrt{6}-9 \sqrt{\epsilon } \phi _0}{3 \sqrt{\epsilon } \phi _0}\big]$ \\ [0.7ex]\hline
  $P_{3\pm}$ & $\Big[\frac{1}{\lambda  \phi _0},\frac{2-3 \lambda  \phi _0}{2 \lambda  \phi _0},\frac{-3 \sqrt{24 \lambda ^2 \epsilon ^2 \phi _0^2-7 \lambda ^4 \epsilon  \phi _0^2}-3 \lambda ^2 \sqrt{\epsilon } \phi _0}{4 \lambda ^2 \sqrt{\epsilon } \phi _0},\frac{3 \sqrt{24 \lambda ^2 \epsilon ^2 \phi _0^2-7 \lambda ^4 \epsilon  \phi _0^2}-3 \lambda ^2 \sqrt{\epsilon } \phi _0}{4 \lambda ^2 \sqrt{\epsilon } \phi _0}\Big]$ \\ [0.7ex]\hline
  $P_4$ & $\big[\frac{\lambda }{3 \epsilon  \phi _0},\frac{\lambda ^2-6 \epsilon }{2 \epsilon },\frac{\lambda ^2-3 \epsilon }{\epsilon },\frac{2 \lambda -3 \lambda ^2 \phi _0}{6 \epsilon  \phi _0}\big]$ \\ [0.7ex]\hline
  $P_5$ & $\Big[\frac{c_1}{c_2},-\frac{c_1}{c_2}-3,\frac{c_1}{c_2}+3,\frac{2 c_1}{c_2}+3\Big]$ \\ [0.7ex]\hline
  $P_{6,7,8}$ & $[0, \Gamma_{6,7,8}, \Delta_{6,7,8}, \Theta_{6,7,8}]$ \\ [0.7ex]\hline
 \end{tabular}
 \caption{The eigenvalues for the critical points of the string inspired teleparallel cosmology.} \label{Table2}
 \end{table}

\begin{table}[H]
\centering
 \centering
\resizebox{17cm}{!}{\begin{tabular}{||c | c | c | c|c| c||} 
 \hline
 Cr.P. & Existence & $\Omega_{\rm m}$ &$w_{\rm eff}$ & Acceleration & Stability  \\ [0.5ex] \hline\hline
 $P_1$ & Always& 1 & 0& Never & Saddle\\ [0.5ex]
 \hline
 \multirow{2}{*}{$P_{2,+}$} &  \multirow{2}{*}{$\epsilon>0$} & \multirow{2}{*}{0}&  \multirow{2}{*}{1} &  \multirow{2}{*}{Never} & Unstable if $(\lambda \leq 0\land \epsilon >0\land \phi_0>0)\lor \left(\lambda >0\land \epsilon >\frac{\lambda ^2}{6}\land \phi_0>0\right)$ \\ [0.5ex] 
 &  & &  &   & Saddle otherwise \\ [0.5ex] \hline
 \multirow{2}{*}{$P_{2,-}$} &  \multirow{2}{*}{$\epsilon>0$}& \multirow{2}{*}{0} &  \multirow{2}{*}{1} &  \multirow{2}{*}{Never} & Unstable if $ (\lambda >0\land \epsilon >0\land \phi_0<0)\lor \left(\lambda \leq 0\land \epsilon >\frac{\lambda ^2}{6}\land \phi_0<0\right)$ \\ [0.5ex] 
 &  &  &   & & Saddle otherwise \\ [0.5ex] \hline
 \multirow{2}{*}{$P_{3,+}$} &  \multirow{2}{*}{$\lambda >0\land \epsilon >0$} &\multirow{2}{*}{$1-\frac{3 \epsilon }{\lambda ^2}$} & \multirow{2}{*}{0} &  \multirow{2}{*}{Never} & Stable if $\lambda >0\land \frac{7 \lambda ^2}{24}\leq \epsilon <\frac{\lambda ^2}{3}\land \phi_0<0$ \\ [0.5ex] 
 &  &  &   & & Saddle otherwise \\ [0.5ex] \hline
  \multirow{2}{*}{$P_{3,-}$} &  \multirow{2}{*}{$\lambda <0\land \epsilon >0$} & \multirow{2}{*}{$1-\frac{3 \epsilon }{\lambda ^2}$} &  \multirow{2}{*}{0} &  \multirow{2}{*}{Never} & Stable if $\lambda <0\land \frac{7 \lambda ^2}{24}\leq \epsilon <\frac{\lambda ^2}{3}\land \phi_0>0$ \\ [0.5ex] 
 &  &  & &  & Saddle otherwise \\ [0.5ex] \hline
 \multirow{3}{*}{$P_{4}$} &  \multirow{3}{*}{$\epsilon <0\lor 6 \epsilon \geq \lambda ^2$} & \multirow{3}{*}{0} &  \multirow{3}{*}{$\frac{\lambda ^2}{3 \epsilon }-1$} &  \multirow{3}{*}{If $\epsilon <0\lor 2 \epsilon >\lambda ^2$} & Stable if: $(\epsilon <0\land 3 \lambda  \phi_0<2\land ((\lambda <0\land \phi_0<0)\lor (\lambda >0\land \phi_0>0)))$ \\ [0.5ex]
 & & & & & $\lor \left(3 \epsilon >\lambda ^2\land ((\phi_0>0\land \lambda <0)\lor (\lambda >0\land \phi_0<0))\right)$ \\ [0.5ex]
 &  &  & &  & Saddle otherwise \\ [0.5ex] \hline
  $P_{5}$ & $c_1\neq0$ & $\frac{2 c_1}{3 c_2}+1$ &  0 &  \multirow{2}{*}{If $\left.c_1<0\land 0<c_2<-\frac{c_1}{2}\right)\|c_1>0\land -\frac{c_1}{2}<c_2<0$ } & Always Saddle  \\ [2ex] \hline
  \multirow{3}{*}{$ P_{6}$} & $\frac{\alpha  z^3 (2 c_1 c_3-3 c_1 c_4+9 c_2 c_3-9 c_2 c_4)-c_3}{c_3 (3 \lambda  \phi_0-1)-3 c_4 \lambda  \phi_0}\geq0$ &\multirow{3}{*}{0}  & \multirow{3}{*}{-1}  & \multirow{3}{*}{Always} &\multirow{3}{*}{See discussion and Fig.~\ref{fig:fig1}}\\ [2ex]
  &and $-\frac{2 \lambda  \phi_0-2 \alpha  z^3 (-c_1\lambda \phi_0+c_1+3 c_2)}{\alpha  z (-3 c_3 \lambda  \phi_0+c_3+3 c_4 \lambda  \phi_0)}\geq0$ & & & & \\ [2ex]
  & and $z (3 c_3 \lambda  \phi_0-c_3-3 c_4 \lambda  \phi_0)\neq0$ & & & & \\ [2ex]\hline
  $P_{7,\pm}$ &$c_1\geq -3 c_2\land c_3>0$ and $2 c_1 c_3-3 c_1 c_4+9 c_2 c_3-9 c_2 c_4\neq0$ & 0 &  -1 &Always  & See discussion and Fig.~\ref{fig:fig2}  \\ [2ex] \hline
    $P_{8,\pm}$ &  $\frac{-c_1-3 c_2}{c_1 (\lambda  \phi_0-1)-3 c_2}\geq 0$ and $c_1 (\lambda  \phi_0-1)-3 c_2\neq0$ & 0 & -1  &  Always& See discussion and Fig.~\ref{fig:fig3}  \\ [2ex] \hline
 \end{tabular}}
 \caption{Physical properties of the critical points for the exponential coupling model.} \label{Table3}
 \end{table}

\begin{figure}
  \centering
    \includegraphics[width=0.3\textwidth]{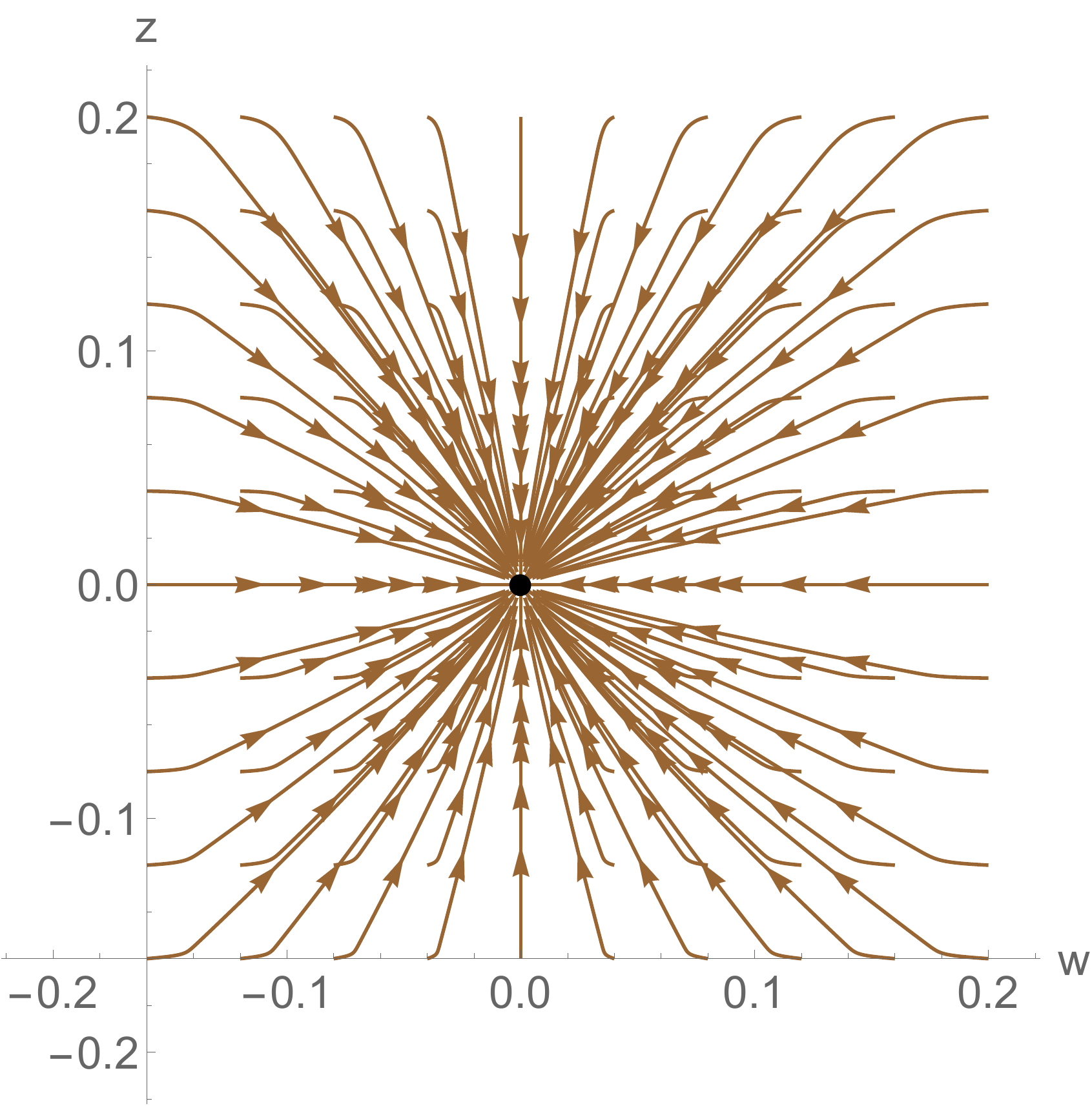}
  \caption{The phase space convergence of the numerical solutions towards $P_4$ critical point in the O--Z--W plane.}
  \label{fig:fig4a}
\end{figure}

\begin{figure}
  \centering
    \includegraphics[width=0.3\textwidth]{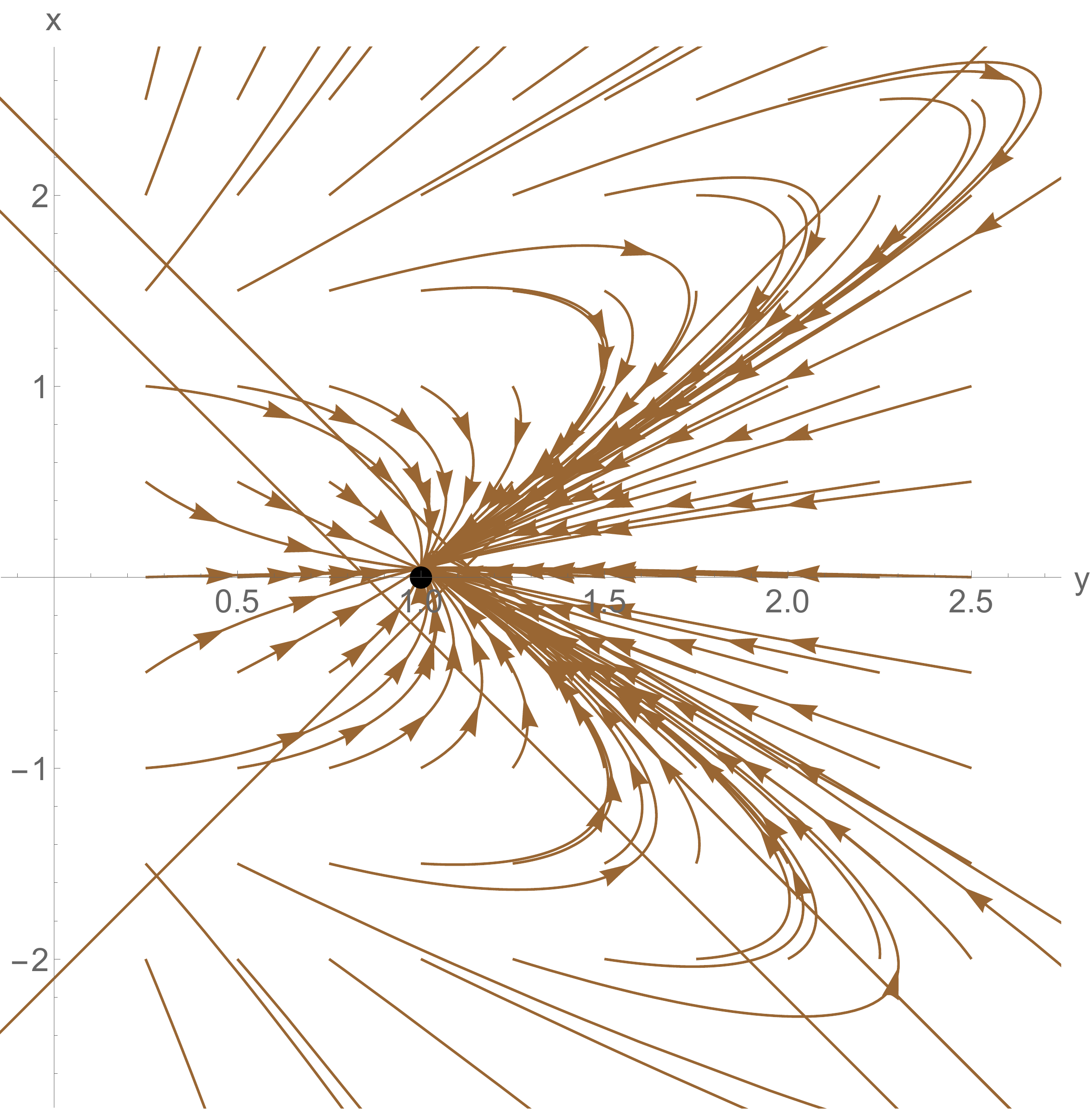}
  \caption{The phase space convergence of the numerical solutions towards $P_4$ critical point in the O--X--Y plane.}
  \label{fig:fig4b}
\end{figure}

\begin{figure}
  \centering
    \includegraphics[width=0.3\linewidth]{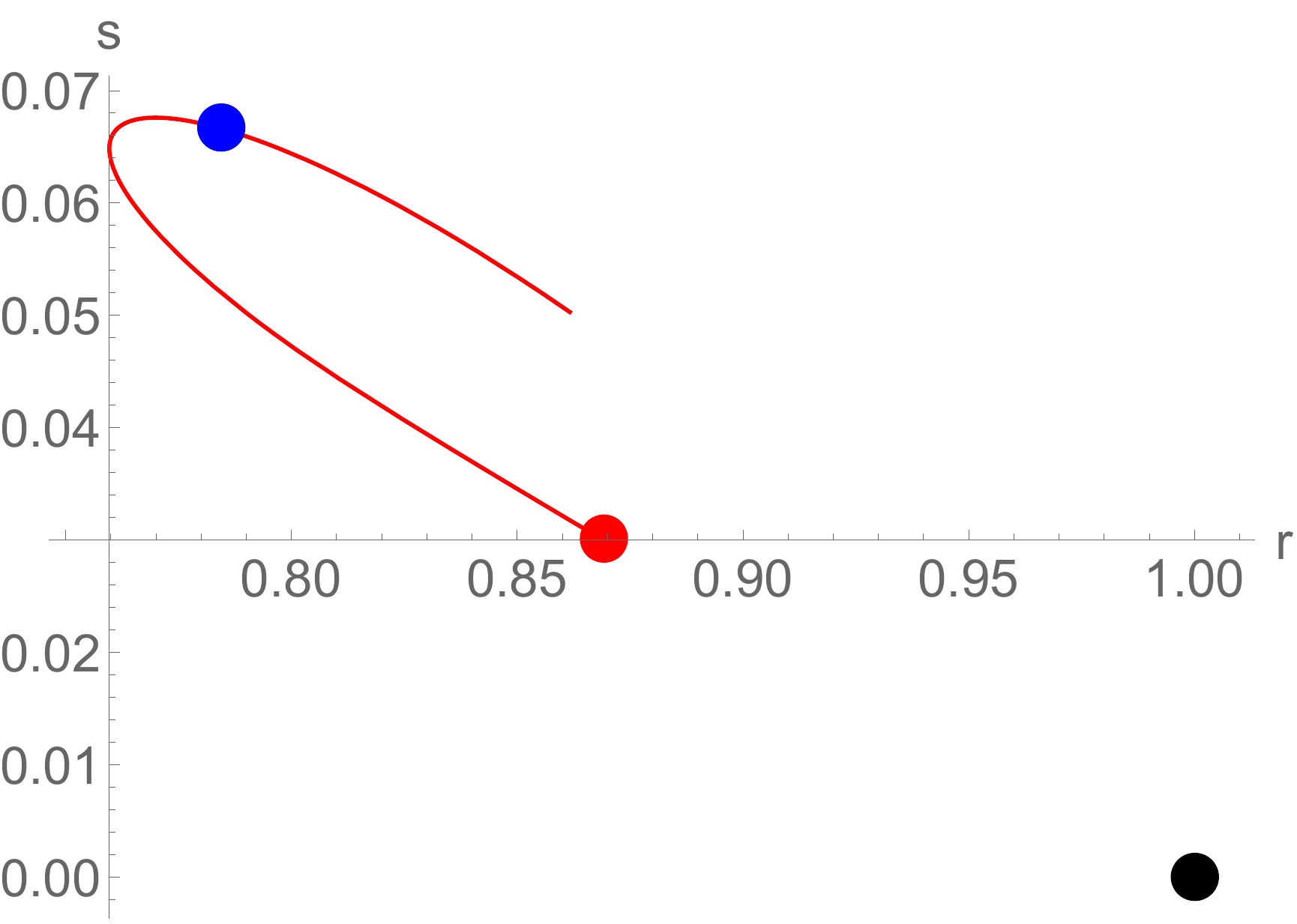}
    \includegraphics[width=0.3\linewidth]{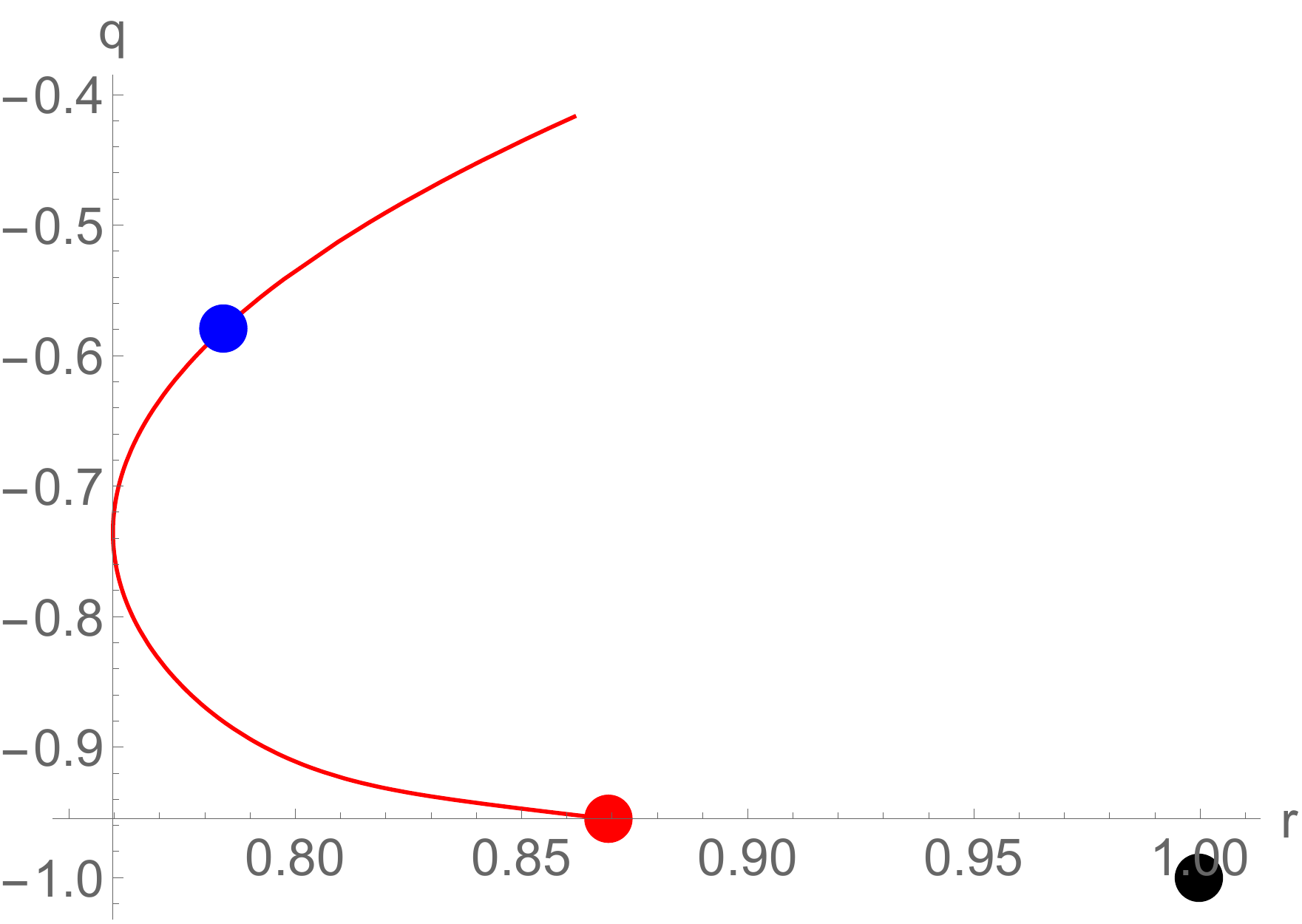}
    \includegraphics[width=0.3\linewidth]{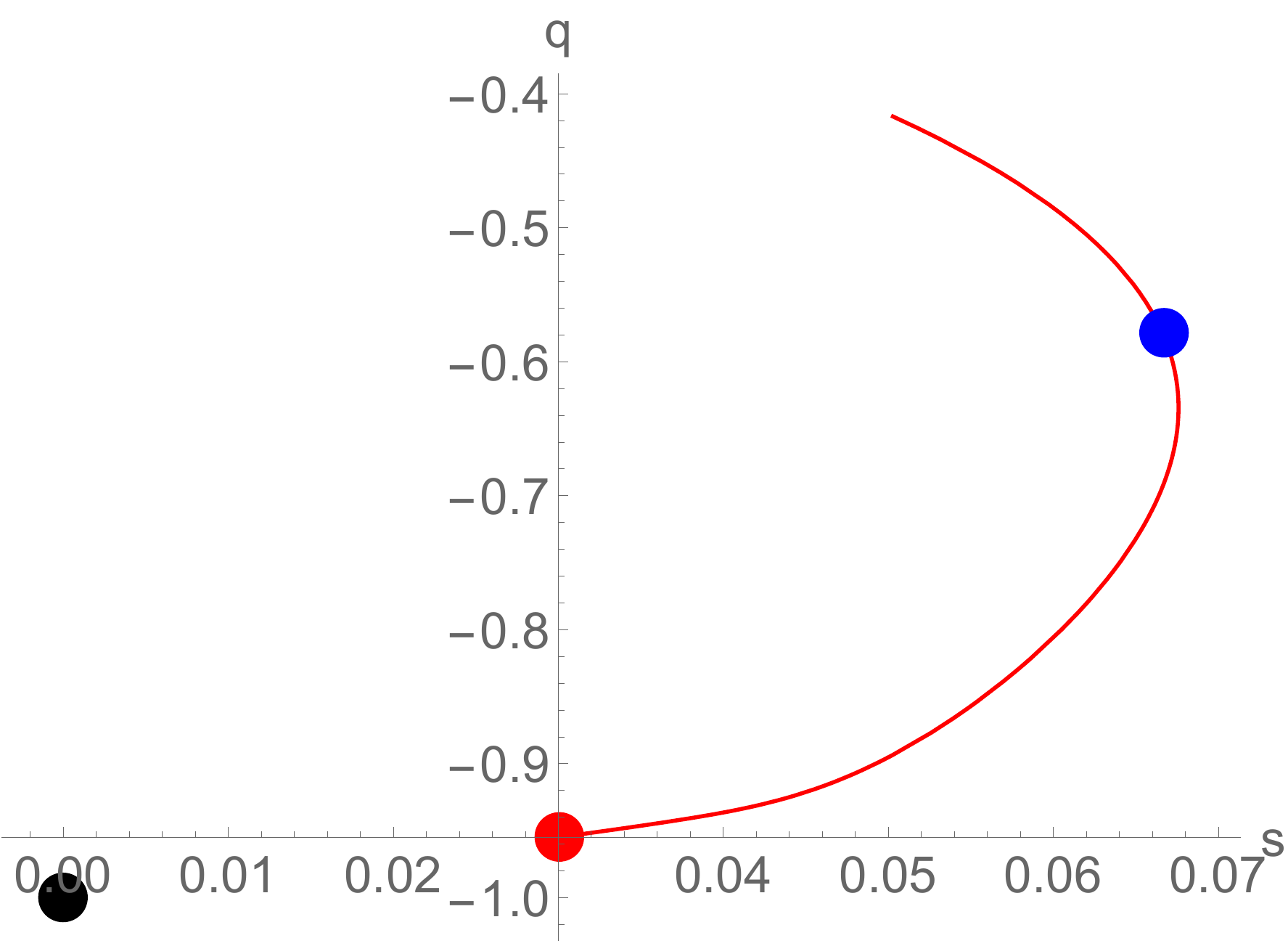}
  \caption{The evolution of the statefinder parameters $\{r,s\},\{r,q\},\{s,q\}$ in the phase space corresponding to the $P_4$ critical point. In the graphs the dot below the horizontal axis of the evolution represents the $\Lambda$CDM epoch. The dot on the horizontal axis in the evolution presents the final value of the parameters corresponding to the $P_4$ solution. Finally, the remaining dots associated to the curve marks the late time values of the statefinder parameters. The following specific conditions have been used: ($\omega=0, \epsilon=+1, \alpha=1, \phi_0=-1, c_1=1, c_2=1, c_3=0, c_4=0, \lambda=0.3, x_0=0.3, y_0=0.9, z_0=0.5, w_0=0.02)$. The initial conditions have been tuned in order to obtain $\Omega_m=0.31$ at the present time.
  }
  \label{fig:fig4c}
\end{figure}

\begin{figure}
  \centering
    \includegraphics[width=0.3\textwidth]{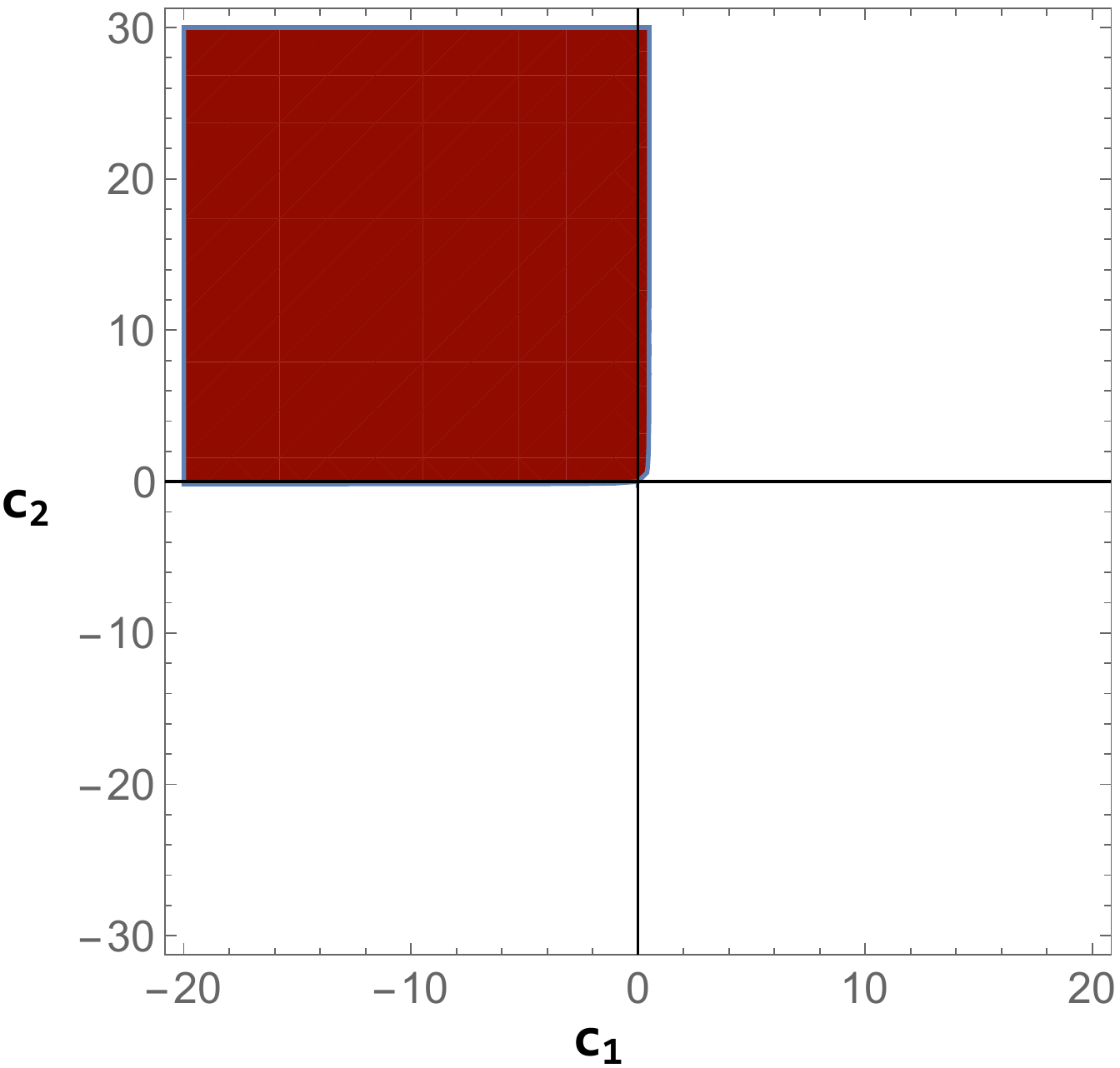}
  \caption{The figure describe a region for the $P_6$ critical point which is associated to a saddle dynamical behavior $(\lambda=1, \phi_0=1, \epsilon=1, \alpha=-1, z=1, c_3=1, c_4=1)$.}
  \label{fig:fig1}
\end{figure}

\begin{figure}
  \centering
    \includegraphics[width=0.3\textwidth]{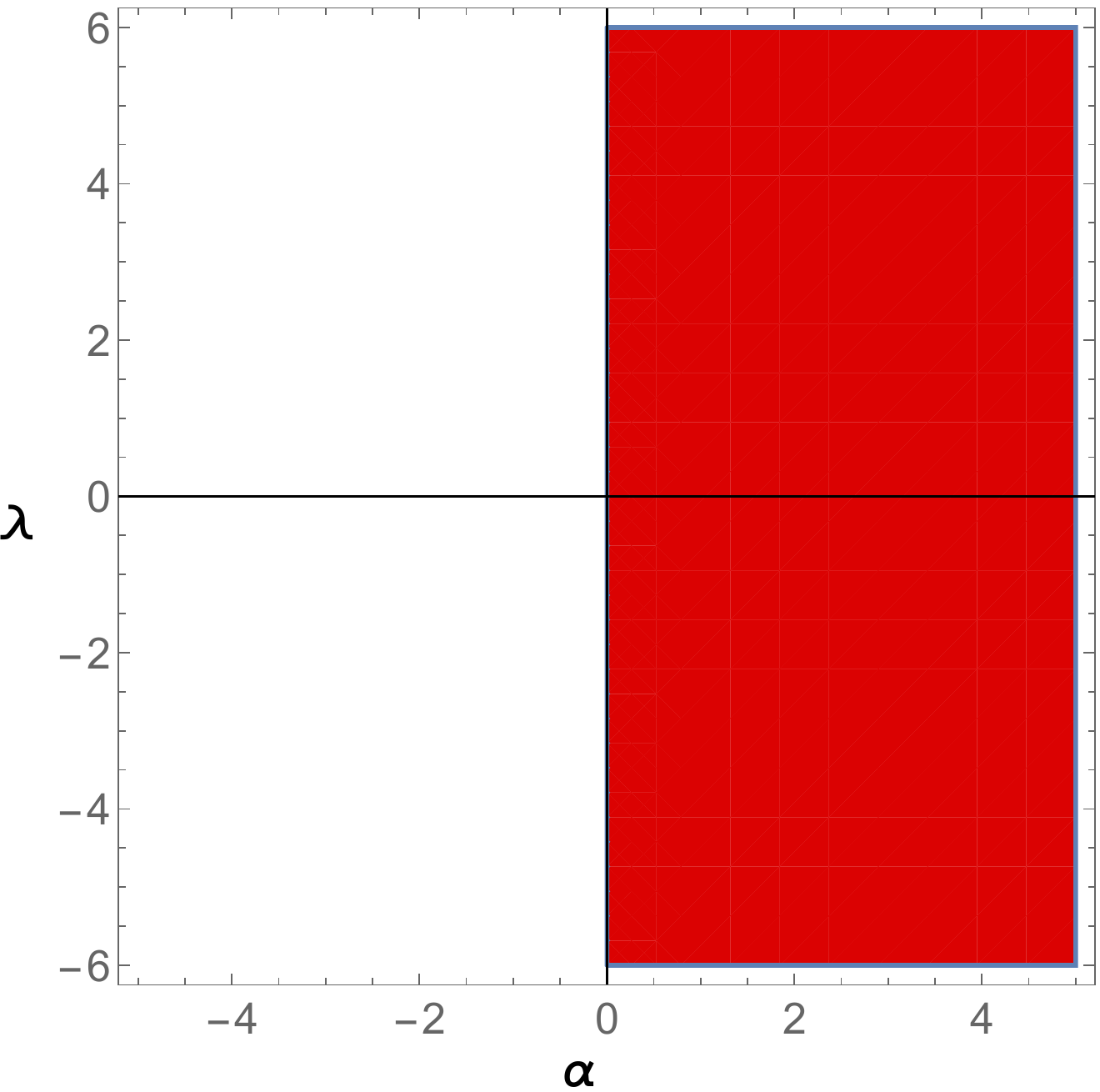}
  \caption{The figure displays a possible saddle interval in the case of the $P_{7+}$ solution $(c_1=6, c_2=0.1, c_3=160, c_4=80, \epsilon=1, \phi_0=1)$.}
  \label{fig:fig2}
\end{figure}

\begin{figure}
  \centering
    \includegraphics[width=0.3\textwidth]{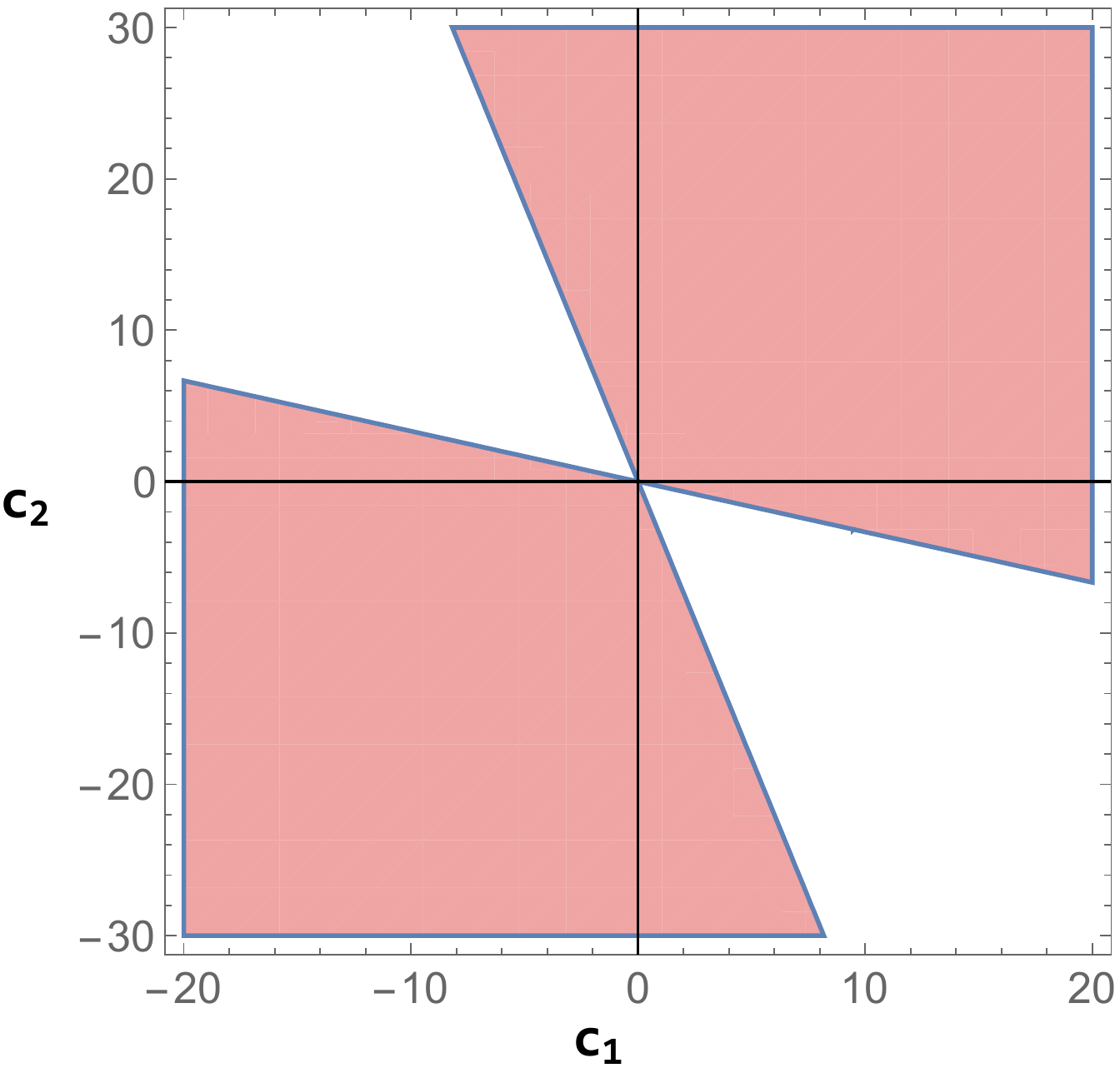}
  \caption{The boundary area associated to the $P_{8+}$ critical point where the dynamical behavior corresponds to a saddle cosmological solution $(\alpha=-10, \epsilon=1, \lambda=1, \phi_0=-10)$.}
  \label{fig:fig3}
\end{figure}

\begin{figure}
  \centering
    \includegraphics[width=0.3\textwidth]{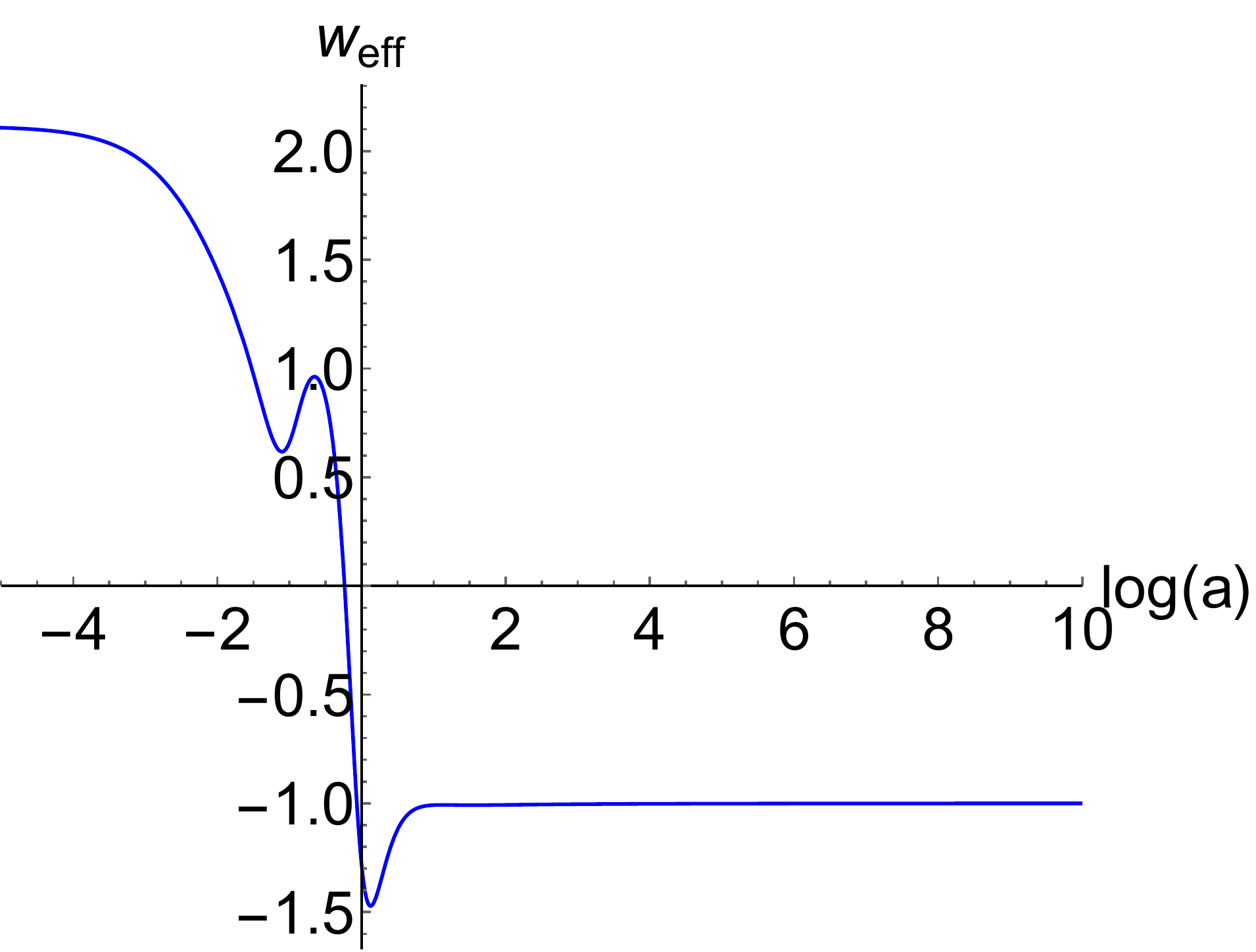}
  \caption{The evolution of the effective equation of state from a super--stiff epoch towards a de--Sitter cosmological era.}
  \label{fig:fig5}
\end{figure}

\par 
In what follows we shall discuss the phase space structure and the physical features of the corresponding critical points, as well as the dynamical effects due to the stability criteria. In the case of exponential potential energy and string inspired couplings the structure of the phase space consists of four classes of critical points. The general features of the obtained critical points are described in Tables~\ref{Table1}--\ref{Table3}, where we have introduced the location in the phase space structure, the corresponding existence conditions and the dynamical properties of the solutions due to the specific expression of the obtained eigenvalues. Note that the eigenvalues for the non--hyperbolic critical points $P_{6,7,8}$ denoted as $\Gamma, \Delta, \Theta$ have complicated expressions and are not displayed in the manuscript. For these specific solutions, the dynamical analysis relies on numerical evaluations of the resulting eigenvalues. The analysis showed that the structure of the phase space is composed of four classes of critical points, corresponding to different cosmological solutions. From the definition of the dimensionless variables the auxiliary variable $x$ is related to the kinetic energy of the scalar field as quintessence or phantom, respectively, $y$ the value of the potential energy, while $z$ and $v$ encodes the effects of the specific value of the scalar field $\phi$. We note that in the case of a zero kinetic energy the scalar field is frozen to a specific value, without a dynamical evolution. The first critical point represents the origin of the phase space, corresponding to a matter-dominated epoch, which is always a saddle cosmological solution. This solution corresponds to the first class of critical points, where the effective equation of state is describing a matter epoch. We note that in the structure of the phase space there are three critical points which belong to the matter epoch, $P_{1}, P_{3}, P_{5}$. For the first critical point $P_{1}$ the solution corresponds to a matter-dominated era, while for the remaining points $P_{3,5}$ the value of the matter density parameter is sensitive to the potential energy strength and the sign of the kinetic energy. All of these critical points represent possible solutions which can describe the matter-dominated epoch of the Universe in the recent past. The second class of cosmological solutions is corresponding to a stiff--fluid dynamics, which appears in the case of $P_{2\pm}$ critical points. This solution is not of great interest in modern cosmology and cannot represent a stable solution, which can be noted from the results of the dynamical analysis in Table~\ref{Table2}. 
\par 
The third class of cosmological solutions is associated with a de--Sitter epoch which appears for the $P_{6,7,8}$ critical points, an era where the dynamics mimic a cosmological constant behavior, where the matter content of the Universe is negligible in terms of density parameters. These cosmological solutions represent critical points which are non--hyperbolic due to the existence of one zero eigenvalue in all of the cases. The rest of the eigenvalues denoted as $[\Gamma_{6,7,8}, \Delta_{6,7,8}, \Theta_{6,7,8}]$ have cumbersome expressions and are not displayed in the manuscript. Due to these reasons, we have analyzed the dynamical features of the corresponding points by considering  numerical estimations, displaying different regions which are connected to a saddle dynamical behavior in Figs.~\ref{fig:fig1}--\ref{fig:fig3}, for specific values of the coupling parameters and constants.
\par 
The last class is represented by the $P_4$ critical point where the kinetic energy of the scalar field and the potential energy are affecting the location in the phase space structure. The matter density parameter is equal to zero, denoting the full domination of the geometrical dark energy density parameter over the matter sector. The effective equation of state is sensitive to the potential energy strength encoded into the $\lambda$ parameter and the sign of the kinetic energy expressed into the $\epsilon$ constant. This solution represents a viable epoch that can explain the accelerated expansion of the Universe and the non--negligible deviation from the $\Lambda$CDM behavior. The evolution towards $P_4$ critical point is presented in Figs.~\ref{fig:fig4a}--\ref{fig:fig4b}. In Table~\ref{Table2} we have obtained specific conditions for the stability of this solution. It can be seen that from a dynamical point of view the stability conditions are affected by the type of the kinetic energy, the potential energy strength and the value of the $\phi_0$ parameter, associated with the string inspired coupling. The deviation from the $\Lambda$CDM model is represented in Fig.~\ref{fig:fig4c} by using the statefinder analysis. In these figures the initial conditions are fined tuned in order to obtain the value $\Omega_m=0.31$ at the present time. In the figures, the black dot represents the $\Lambda$CDM cosmological epoch, while the red one presents the final value of the statefinder parameters corresponding to the $P_4$ solution. Finally, in the graphs, the blue dot marks the late time values of the statefinder parameters $\{r,s,q\}$. We can note that the presented numerical solution has a non--negligible deviation from the $\Lambda$CDM cosmological epoch at late times. In Fig.~\ref{fig:fig5} we have represented the evolution of the effective equation of state in this model from an initial stage corresponding to a super--stiff fluid scenario. It is observed that the total equation of state has a de--Sitter behavior at the final times, crossing the phantom divide line in the late time evolution.

\section{Numerical features of the model}\label{sec:numerics}

\subsection{Torsion coupling with a scalar field}
In the following we shall consider a specific model which takes into account only the torsion coupling for the scalar field:
\begin{equation}\label{model1}
F_{1}(\phi)=c_{1}\alpha e^{\phi/\phi_{0}}\,, ~~~~~F_{2}(\phi)=F_{3}(\phi)=F_{4}(\phi)=0\,.
\end{equation}
Considering that the potential energy is represented as an exponential function described in the Eq.~\eqref{potentialchoice}, we get the following form for the energy density and pressure of the scalar field,
\begin{equation}
\rho_{\phi}=\frac{1}{2}\epsilon \dot{\phi}^{2}+V(\phi)-3c_{1}\alpha H^{2}e^{\phi/\phi_{0}}\,,
\end{equation}
\begin{equation}
p_{\phi}=\frac{1}{2}\epsilon \dot{\phi}^{2}-V(\phi)+\left(3H^{2}+2\dot{H}\right)c_{1}\alpha e^{\phi/\phi_{0}}+\frac{2c_{1}\alpha}{\phi_{0}}\dot{\phi}He^{\phi/\phi_{0}}\,.
\end{equation}
Then, the Klein-Gordon equation becomes,
\begin{equation}
\epsilon \ddot{\phi}+3\epsilon H\dot{\phi}+V'(\phi)+\frac{3c_{1}\alpha}{\phi_{0}}H^{2}e^{\phi/\phi_{0}}=0\,.
\end{equation}
By varying the value of $\epsilon$ we can study the numerical features of the cosmological model both for the canonical ($\epsilon=1$) and phantom ($\epsilon=-1$) scalar fields. Due to simplicity we shall consider only the quintessence case, where $\epsilon=+1$.
We can further define the dark energy equation of state as given by:
\begin{equation}
w_{\rm de}=\frac{p_{\phi}}{\rho_{\phi}}\,.
\end{equation}
Next, we will use the following expression for the Friedmann acceleration equation,
\begin{equation}\label{accl}
\frac{\ddot{a}}{a}=-\frac{1}{6}\left(\rho_{\phi}+3p_{\phi}\right)-\frac{\Omega_{m0}H_{0}^{2}}{2a^{3}}\,,
\end{equation}
taking into account the following values which characterize the present time, $\Omega_{m0}\sim 1-\Omega_{de}\sim 0.30,~t_{0}\sim 0.96,~H_{0}\sim a_{0}\sim 1$. Moreover, in order to simplify the evolutionary aspects, we shall consider that at the initial stage we can describe the evolution of the scale factor with the following relation
\begin{equation}
a(t)\sim \left(\frac{9\Omega_{m0}}{4}\right)^{1/3}t^{2/3}\,,
\end{equation}
a relation specific for a matter-dominated epoch by neglecting any coupling function of the scalar field. 

\begin{figure}[h!]
  \centering
    \includegraphics[height=0.2\textheight]{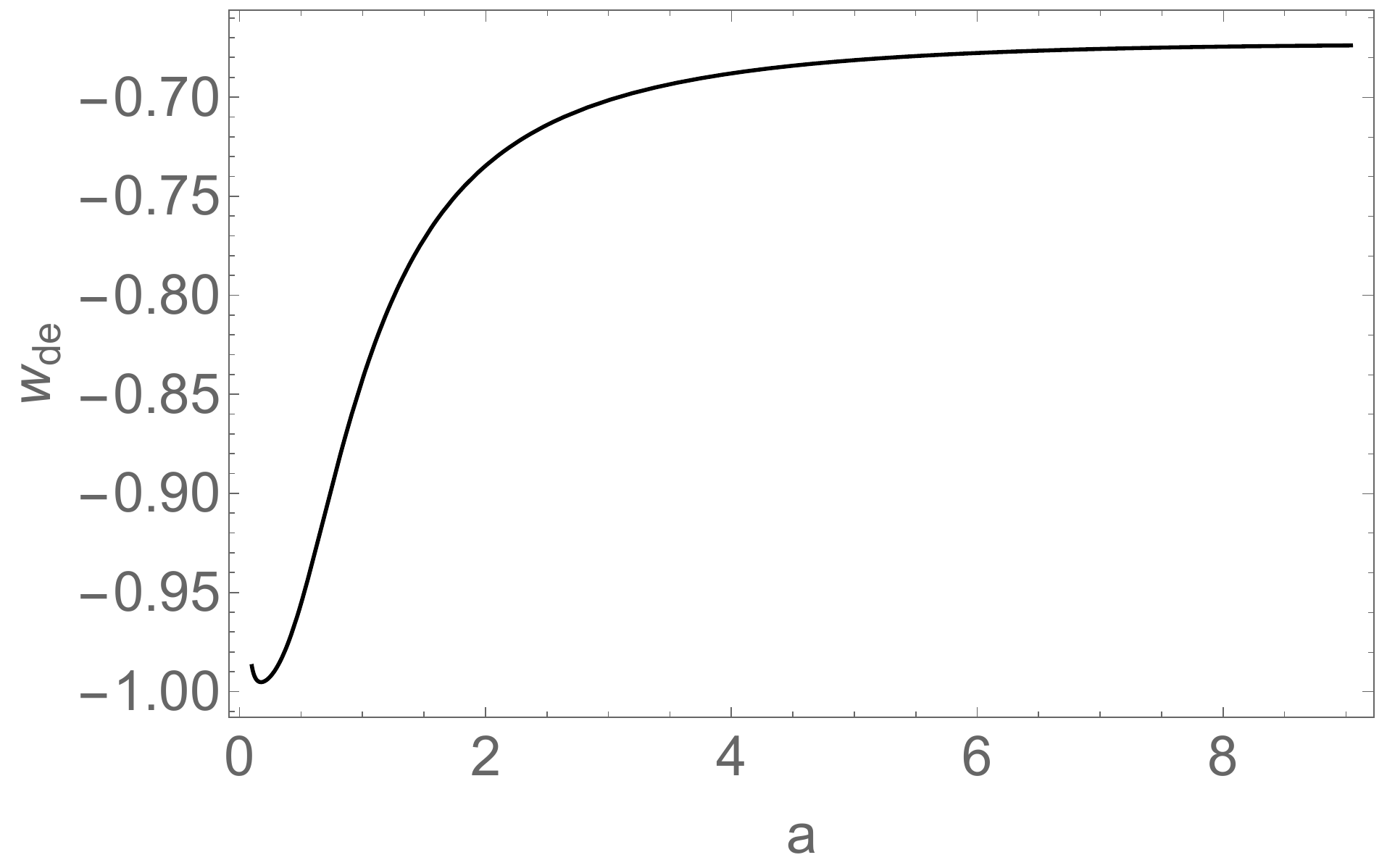}
    \includegraphics[height=0.2\textheight]{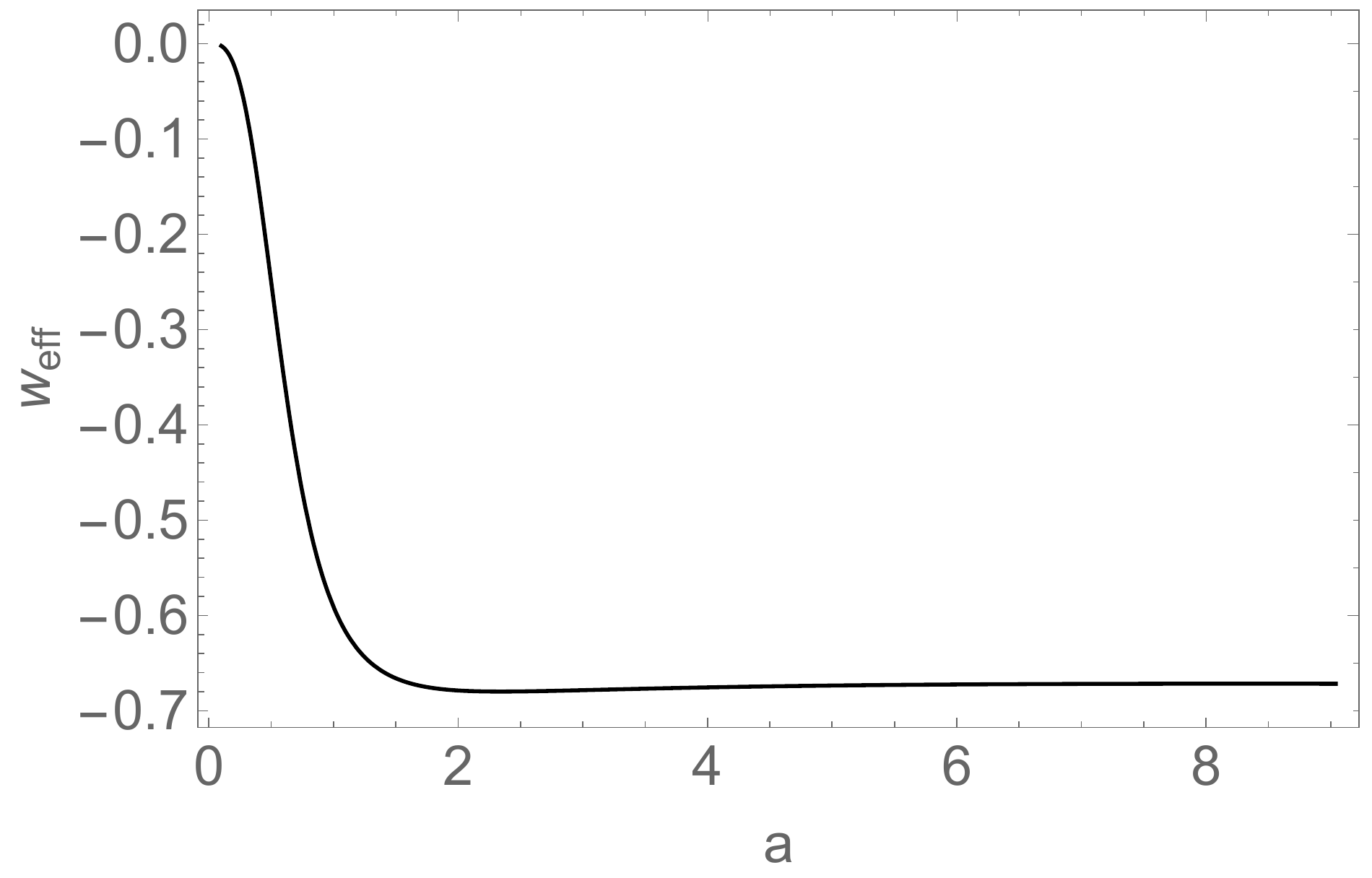}
  \caption{The evolution of the dark energy equation of state $w_{\rm de}$ and the total equation of state $w_{\rm eff}$  for the torsion coupling model in terms of the cosmic scale factor. $[c_1=-0.2, \lambda=1, V_0=0.082, \phi_0=2, \alpha=0.2] $ }
  \label{fig:figt1}
\end{figure}

This method has been used in a variety of recent papers, see for example Refs. \cite{Bahamonde:2018miw,Lykkas:2015kls,Perivolaropoulos:2004yr}. The results of our analysis are displayed in Fig.~\ref{fig:figt1}, where we have displayed a possible evolution for the dark energy equation of state $w_{\rm de}$ in terms of the cosmic scale factor $a$, considering the fine--tuning for the initial conditions. We have also presented the evolution of the total equation of state from deep into the matter era. The figure shows that at the beginning of the matter-dominated epoch the scalar field model evolves near the cosmological constant, while at late times the evolution corresponds to a  quintessence regime.

\subsection{Boundary coupling with a scalar field}
In this section we consider a specific model by taking into account only the boundary coupling for the scalar field, where:
\begin{equation}\label{model2}
F_{2}(\phi)=c_{2}\alpha e^{\phi/\phi_{0}}\,, ~~~~~F_{1}(\phi)=F_{3}(\phi)=F_{4}(\phi)=0\,.
\end{equation}
Hence we get the following form for the energy density and pressure of the scalar field,
\begin{equation}
\rho_{\phi}=\frac{1}{2}\epsilon \dot{\phi}^{2}+V(\phi)+\frac{3c_{2}\alpha}{\phi_{0}} H\dot{\phi}e^{\phi/\phi_{0}}\,,
\end{equation}
\begin{equation}
p_{\phi}=\frac{1}{2}\epsilon \dot{\phi}^{2}-V(\phi)-\frac{c_{2}\alpha}{\phi_{0}}\ddot{\phi} e^{\phi/\phi_{0}}-\frac{2c_{2}\alpha}{\phi_{0}^{2}}\dot{\phi}^{2}e^{\phi/\phi_{0}}\,.
\end{equation}
In this case the The Klein-Gordon equation becomes,
\begin{equation}
\epsilon \ddot{\phi}+3\epsilon H\dot{\phi}+V'(\phi)+\frac{3c_{2}\alpha}{\phi_{0}}\left(3H^{2}+\dot{H}\right)e^{\phi/\phi_{0}}=0\,.
\end{equation}

\begin{figure}[h!]
  \centering
    \includegraphics[height=0.2\textheight]{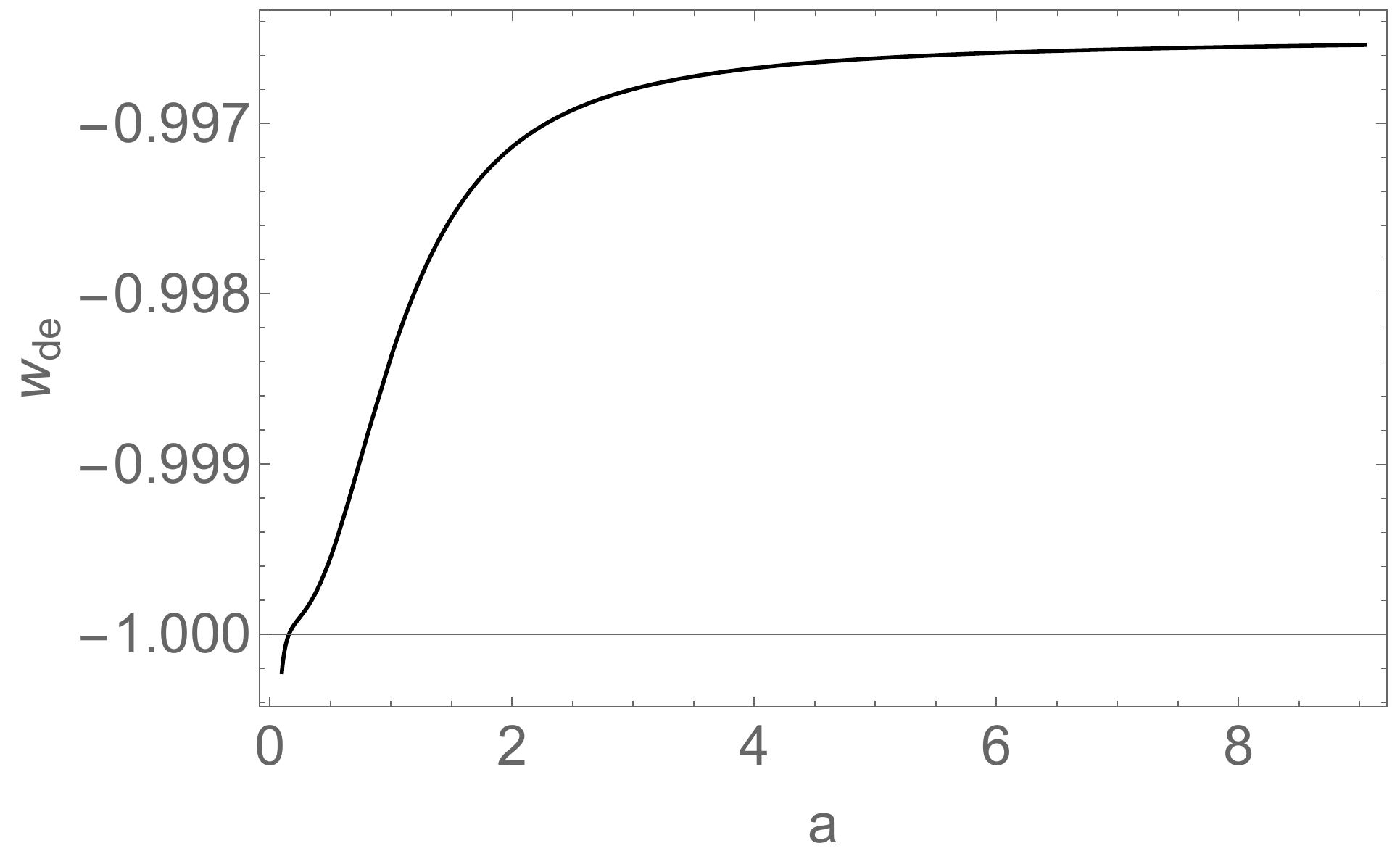}
     \includegraphics[height=0.2\textheight]{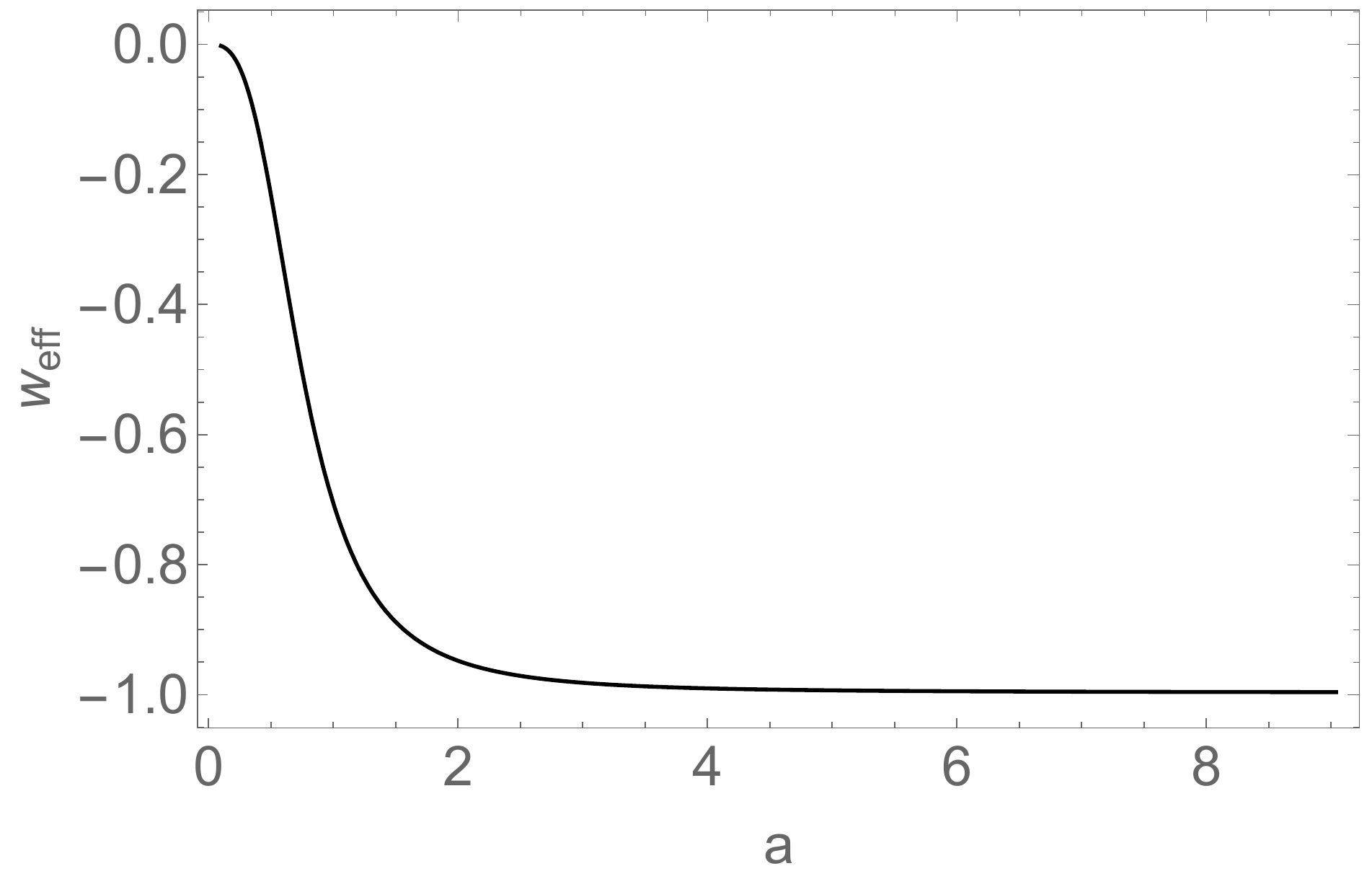}
  \caption{The evolution of the dark energy equation of state $w_{\rm de}$ and the total equation of state $w_{\rm eff}$ for the boundary coupling model in terms of the cosmic scale factor. $[c_2=-0.1, \lambda=0.1, V_0=1.8, \phi_0=1.2, \alpha=1] $ }
  \label{fig:figb1}
\end{figure}

In Fig.~\ref{fig:figb1} we have displayed a possible evolution for the equations of state described by $w_{\rm de}, w_{\rm eff}$ in terms of the cosmic scale factor, considering the fine--tuning for the initial conditions. The figure shows the crossing of the phantom divide line at the beginning of the evolution, near the matter-dominated epoch. At late times the model mimics an evolution near the phantom divide line, acting as a cosmological constant, and it is a cosmological model with reduced variations near the specific boundary.  

\subsection{Gauss-Bonnet coupling with a scalar field}

\subsubsection{Case 1: $T_{G_{3}}$ coupled to $\phi$}
In this case we consider the model where
\begin{equation}\label{model3}
F_{3}(\phi)=c_{3}\alpha e^{\phi/\phi_{0}}\,, ~~~~~F_{1}(\phi)=F_{2}(\phi)=F_{4}(\phi)=0\,,
\end{equation}
described by the following energy density and pressure of the scalar field,
\begin{equation}
\rho_{\phi}=\frac{1}{2}\epsilon \dot{\phi}^{2}+V(\phi)+36c_{3}\alpha H^{4}e^{\phi/\phi_{0}}\,,
\end{equation}
\begin{equation}
p_{\phi}=\frac{1}{2}\epsilon \dot{\phi}^{2}-V(\phi)-\frac{16c_{3}\alpha}{\phi_{0}}\dot{\phi} H^{3}e^{\phi/\phi_{0}}-12c_{3}\alpha H^{2}\left(3H^{2}+4\dot{H}\right)e^{\phi/\phi_{0}}\,.
\end{equation}
Then, the Klein-Gordon equation becomes,
\begin{equation}
\epsilon \ddot{\phi}+3\epsilon H\dot{\phi}+V'(\phi)-\frac{12c_{3}\alpha}{\phi_{0}}H^{4}e^{\phi/\phi_{0}}=0\,.
\end{equation}

\begin{figure}[h!]
  \centering
    \includegraphics[height=0.2\textheight]{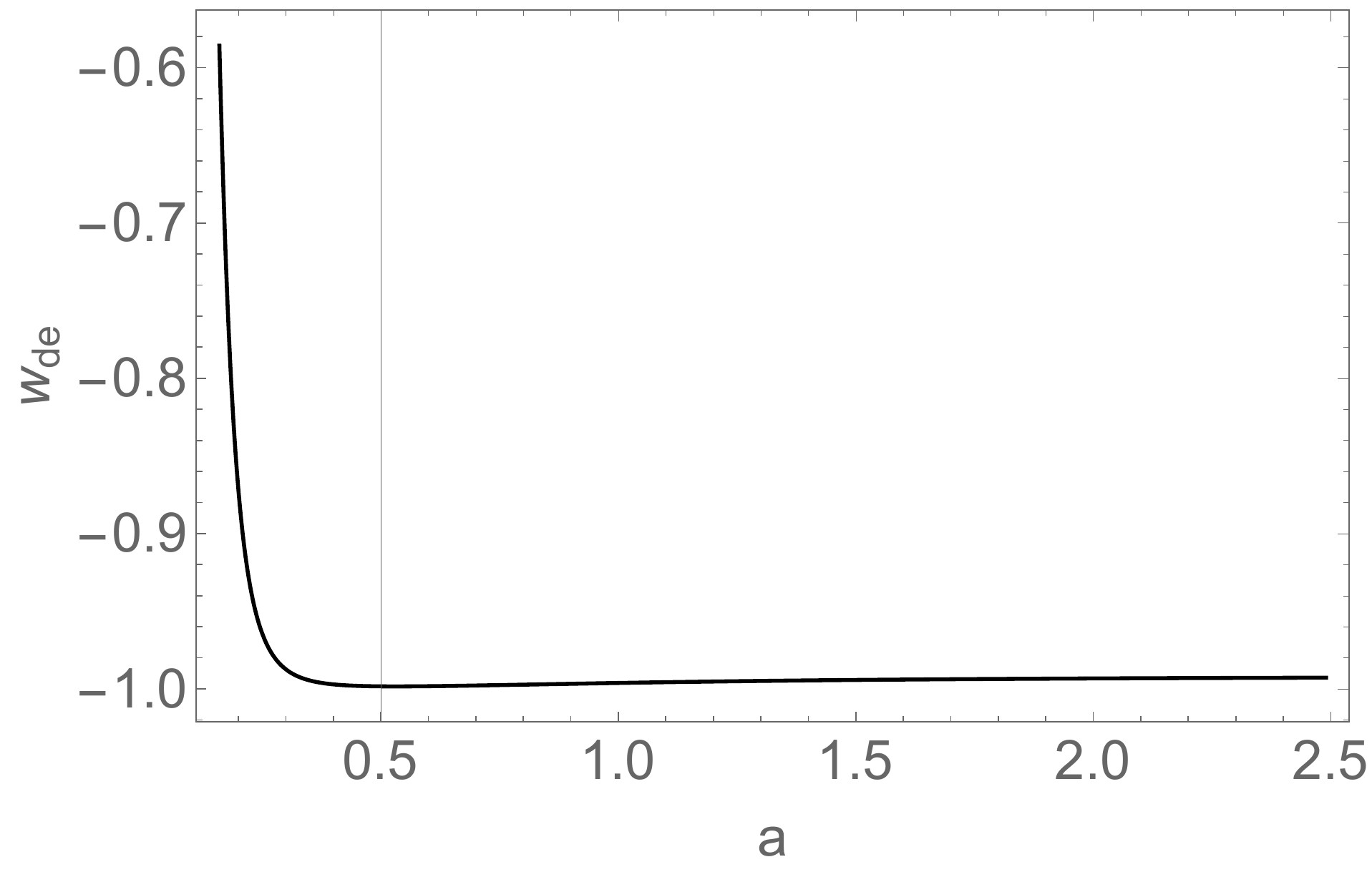}
     \includegraphics[height=0.2\textheight]{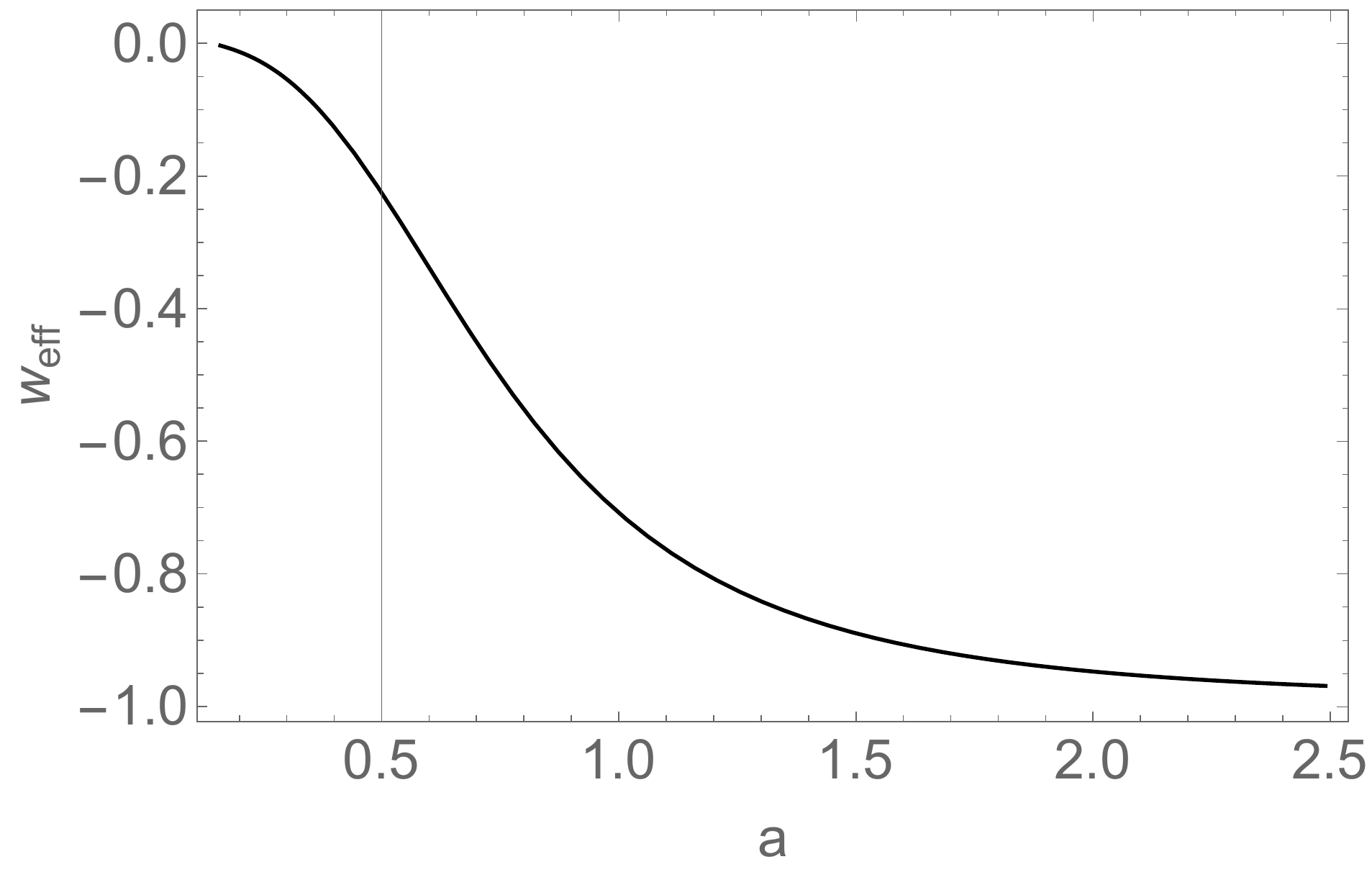}
  \caption{The evolution of the dark energy equation of state $w_{\rm de}$ and the total equation of state $w_{\rm eff}$ for the $T_{G_3}$ coupling model in terms of the cosmic scale factor. $[c_3=1.5, \lambda=0.155, V_0=0.792, \phi_0=2, \alpha=0.7] $}
  \label{fig:figtg3}
\end{figure}

In Fig.~\ref{fig:figtg3} we have displayed a possible evolution for the total effective equation of state in terms of the cosmic scale factor, considering the fine--tuning of the initial conditions. The figure shows the evolution from the matter-dominated epoch towards a dark era where the effective equation of state has a variation near the cosmological constant boundary. In this specific case, the evolution of the scalar corresponds to a quintessence regime, attaining the de--Sitter behavior asymptotically at late times.

\subsubsection{Case 2: $T_{G_{4}}$ coupled to $\phi$}
Here we consider the model characterized by the following values of the coupling functions:
\begin{equation}\label{model4}
F_{4}(\phi)=c_{4}\alpha e^{\phi/\phi_{0}}\,, ~~~~~F_{1}(\phi)=F_{2}(\phi)=F_{3}(\phi)=0\,,
\end{equation}
obtaining the following expressions for the energy density and pressure of the scalar field,
\begin{equation}
\rho_{\phi}=\frac{1}{2}\epsilon \dot{\phi}^{2}+V(\phi)-36c_{4}\alpha H^{4}e^{\phi/\phi_{0}}-\frac{12c_{4}\alpha}{\phi_{0}}\dot{\phi}H^{3}e^{\phi/\phi_{0}}\,,
\end{equation}
\begin{equation}
p_{\phi}=\frac{1}{2}\epsilon \dot{\phi}^{2}-V(\phi)+\frac{4c_{4}\alpha}{\phi_{0}^{2}}\dot{\phi}^{2} H^{2}e^{\phi/\phi_{0}}+\frac{c_{4}\alpha}{\phi_{0}} \left(24H^{3}\dot{\phi}+4H^{2}\ddot{\phi}+8H\dot{H}\dot{\phi}\right)e^{\phi/\phi_{0}}+12H^{2}\left(3H^{2}+4\dot{H}\right)c_{4}\alpha e^{\phi/\phi_{0}}\,.
\end{equation}
In this case the evolution of the scalar field is described by the  Klein-Gordon equation which becomes
\begin{equation}
\epsilon \ddot{\phi}+3\epsilon H\dot{\phi}+V'(\phi)-\frac{12c_{4}\alpha}{\phi_{0}}\dot{H}H^{2}e^{\phi/\phi_{0}}=0\,.
\end{equation}

\begin{figure}[h!]
  \centering
    \includegraphics[height=0.2\textheight]{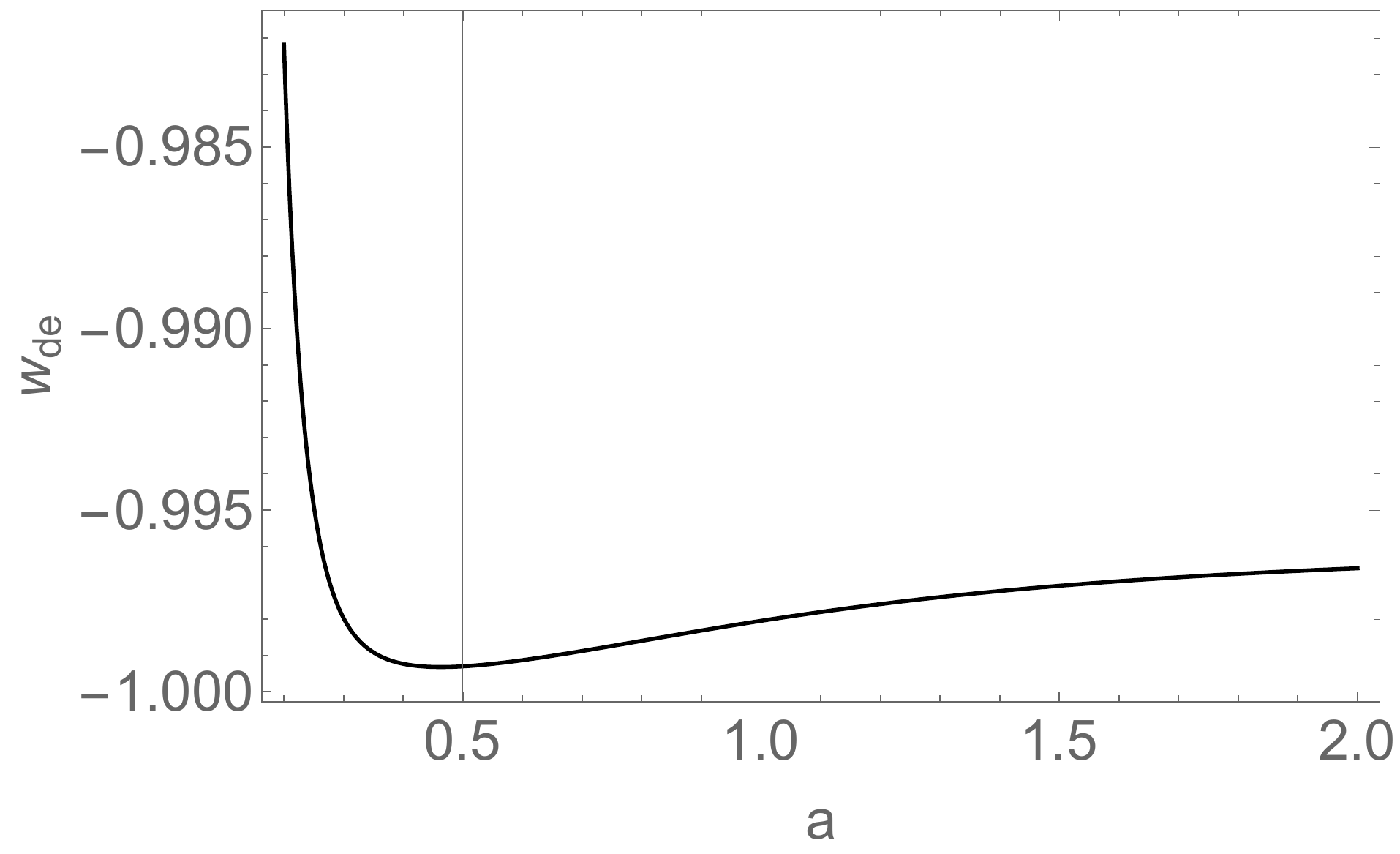}
     \includegraphics[height=0.2\textheight]{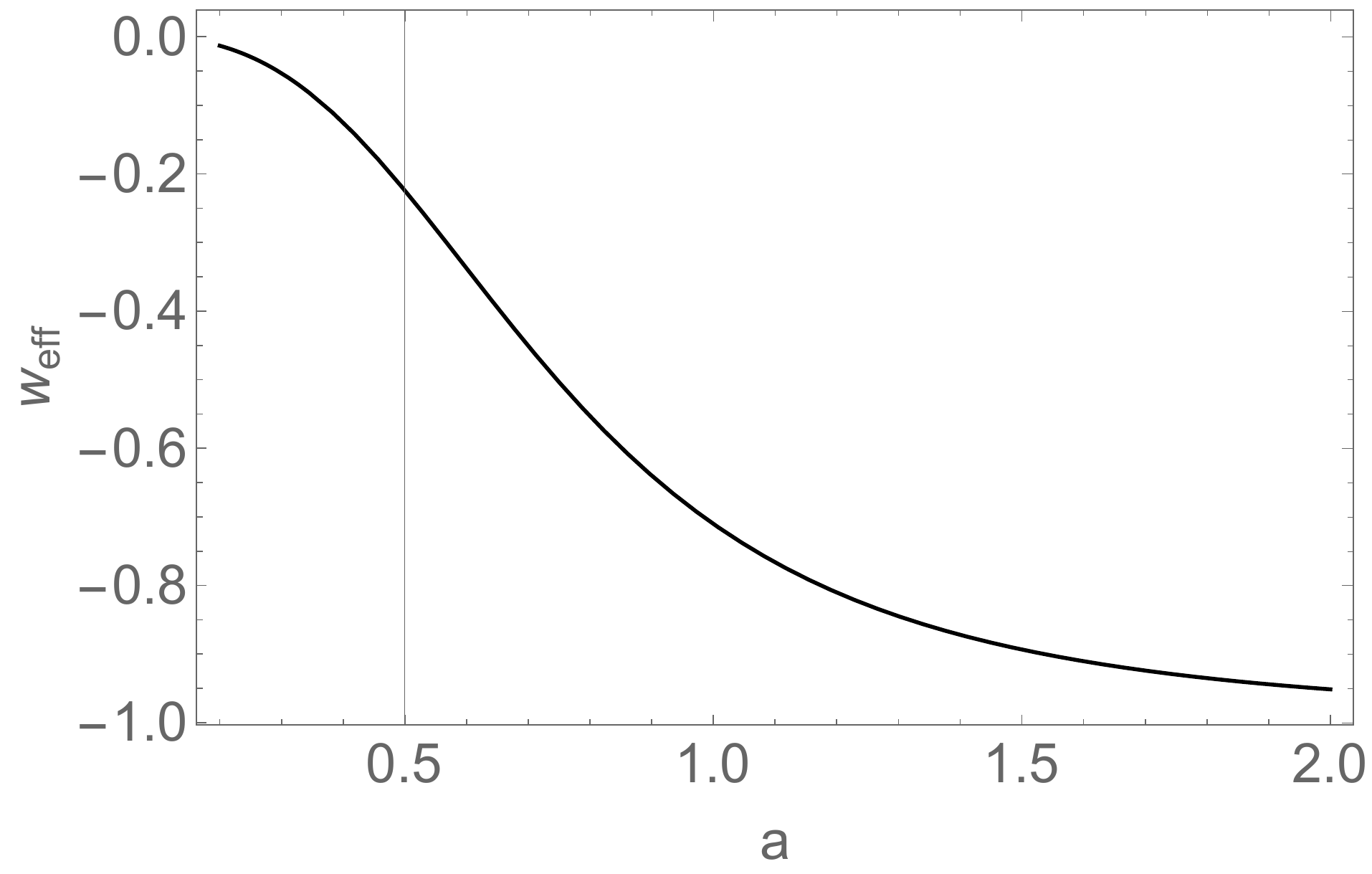}
  \caption{The evolution of the dark energy equation of state $w_{\rm de}$ and the total equation of state $w_{\rm eff}$ for the $T_{G_4}$ coupling model in terms of the cosmic scale factor. $[c_4=-1, \lambda=0.11, V_0=1.1, \phi_0=1.6, \alpha=1] $ }
  \label{fig:figtg4}
\end{figure}

Lastly, in Fig.~\ref{fig:figtg4} we have presented a specific evolution for the total effective equation of state in terms of the cosmic scale factor. The figure also shows the evolution of the dark energy equation of state in the case of a quintessence model coupled to the Gauss--Bonnet $T_{G_4}$ term. As in the previous cases, the dynamics start deep in the matter era, evolving towards an epoch characterized by the domination of the scalar field, acting asymptotically as a cosmological constant at late times.

\section{Summary and Conclusions}
\label{sec:Conclusions}
In this paper, we have studied a generic theory of gravitation in the framework of teleparallel alternative to general relativity. In this case, we have proposed an action based on a general functional which encodes dynamical effects from various invariant terms, including the torsion and boundary terms, as well as the Gauss--Bonnet terms corresponding to the teleparalel gravity framework. After deducing the field equations in a dynamical background corresponding to the FLRW cosmological model, we have obtained the specific form of the modified Friedmann relations for this type of tetrad which reduces the antisymmetric part of the resulting field equations to zero. After encoding the corresponding exponential dependence of the coupling functions specific to string inspired actions, we have analyzed also the physical features of the cosmological model by performing different analytical and numerical techniques. \\

For the analytical investigation, we have used the linear stability methods, transforming the evolution equations to an autonomous system of differential equations, by assuming the corresponding dimensionless variables. In this case, the dimension of the phase space reduces to four independent variables due to the Friedmann constrain equations. For an exponential potential function and string inspired couplings we have determined the structure of the phase space and the location of the corresponding critical points involved, as well as the existing conditions and the resulting dynamical features. From a physical point of view we have obtained four classes of cosmological solutions. The first cosmological solutions correspond to a matter-dominated epoch, a transient era with a saddle dynamical behavior, independent of the values of the coupling parameters and constants. The second cosmological solutions are represented by stiff--fluid solutions which are not interesting from an astrophysical point of view. The third case is associated with a de--Sitter epoch, where the effective equation of state evolves as a cosmological constant, an era which can explain the accelerated expansion of the current Universe. Lastly, the remaining class of dynamical solutions is represented by an epoch where the resulting effective equation of state is sensitive to the values of some of the parameters, representing a possible era that might be associated with an accelerated expansion in some specific cases. Hence, for the specific stringy case, the dynamical investigation shows that the phase space is rich in dynamical solutions and physically viable at the level of background dynamics. From the analytical point of view, we have obtained relations that constrain the model, which can be complemented by observational data analysis in a future work. Such an analysis will give numerical bounds to the model parameters using the recent cosmological data. Finally, it should be stated that this toy model of the universe is a highly generic one with lots of parameters to play around. This is the striking feature of the model. We believe that by proper fine tuning we should be able to realize any desired cosmological scenario. Here we have kept it in its most general form without considering any particular model of interest. Such sub-models can be studied in future works to understand the efficiency of this model to represent the universe. 

As a final approach, we have studied four particular models using a numerical approach. In all of these models, we have assumed an exponential type of coupling, which is inspired by the string-curvature based action. By fine-tunning of the initial conditions, we have found that in all of these models, it is possible to generically describe the evolution of the observed Universe, starting from a matter-dominated era and then, passing through an accelerated expansion of the Universe with the scalar field coupled with either the torsion scalar, the boundary term, or the Gauss-Bonnet invariants.

To conclude our paper, let us briefly describe some properties regarding our theory. The action~\eqref{action} contains a function which depends on 10 scalars that can be mapped to different modified theories of gravity. This theory can be seen as a generalisation of the theory studied in~\cite{Bahamonde:2016kba} by including a scalar field, a kinetic term and also decomposing the Teleparallel Gauss-Bonnet term in all its possible decomposition. The main motivation to study such a theory is related to the well-known low-energy effective string theory theory (which relies on a manifold with curvature and zero torsion - Levi-Civita connection), which has the following action~\cite{Bento:1988sr,Elizalde:2007pi},
\begin{equation}
    \label{eq:action} {\cal S} = \int d^D x
\sqrt{-g}\left[\frac{\mathring{R}}{2\kappa^2}+2X-2V(\phi)+c\alpha'\mathring{G}e^{\frac{\phi}{\phi_{_0}}}+\ldots\right]\,,
\end{equation}
which also contains higher order string corrections. In our case we worked in $D=4$ and the above action can be recovered by setting the function $f$ as~\eqref{f} with the coupling functions being of the form~\eqref{F1}, and then choosing $c_1=c_2=0$ and $c_3=c_4=c$. Clearly, the Teleparallel version that we studied in this paper is more general than the above action. It should be noted that since it is is possible to write down an equivalent version of the stringy effective action, one can conjecture that string theory might be formulated using Teleparallel theory from the very beginning. Furthermore, due to the nature of the torsion tensor containing only up to first derivatives of the tetrad, the string Teleparallel effective action would have more terms than the standard one. Then, this theory might be even richer in this framework. To the best of our knowledge, this has not been studied in full detail in Teleparallel gravity. The above action has more higher terms, and the next one can be written as
\begin{eqnarray}
\mathring{\mathcal{L}}_{c}^{(2)}=c_2\alpha'^2e^{2\frac{\phi}{\phi_{_0}}}\Big(2c_3 \epsilon^{\mu \nu \rho \sigma
\tau \eta} \epsilon_{\mu '\nu '\rho '\sigma ' \tau '\eta '}
\mathring{R}_{\mu\nu}^{\ \ \ \mu'\nu'} \mathring{R}_{\rho\sigma}^{\ \ \ \rho'\sigma'}
\mathring{R}_{\tau\eta}^{\ \ \ \tau'\eta'} + \mathring{R}^{\lambda}{}_{\mu\sigma\nu}
\mathring{R}^{\sigma\nu}{}_{\alpha\beta}\mathring{R}^{\alpha\beta}{}_{\lambda}{}^{\mu}\Big)\,,\label{second2}
\end{eqnarray}
where bar denotes that the quantities are computed with respect to the Levi-Civita connection. The analogue version of the above term has not been derived in Teleparallel yet, but as the standard Gauss-Bonnet term appears in the Teleparallel string action, it is obvious to believe that this term also would appear when one considers a string Teleparallel effective approach. This term can be transform into the Teleparallel language by replacing the Riemannian curvature
\begin{equation}
    \mathring{R}{}^{\lambda}\,_{\mu\sigma\nu} = \mathring{\nabla}_{\nu}K_{\sigma}{}^{\lambda}{}_{\mu} - 
  \mathring{\nabla}_{\sigma}K_{\nu}{}^{\lambda}{}_{\mu} +
  K_{\sigma}{}^{\rho}{}_{\mu}K_{\nu}{}^{\lambda}{}_{\rho} -
  K_{\sigma}{}^{\lambda}{}_{\rho}K_{\nu}{}^{\rho}{}_{\mu} \,,
  \label{relation}
\end{equation}
where $K_{\mu}{}^{\lambda}{}_{\nu}=(1/2)(T^{\lambda}{}_{\mu\nu} - T_{\nu\mu}{}^{\lambda} +T_{\mu}{}^{\lambda}{}_{\nu})$ is the contortion tensor which, clearly is antisymmetric in its last two indices, giving 
\begin{eqnarray}
\mathcal{L}_{c}^{(2)}&=&c_2\alpha'^2e^{2\frac{\phi}{\phi_{_0}}}\Big[2c_3 \epsilon^{\mu \nu \rho \sigma
\tau \eta} \epsilon_{\mu '\nu '\rho '\sigma ' \tau '\eta '}
\Big(K^{\mu'\alpha}{}_\nu K^{\nu'}{}_{\mu\alpha}-K^{\mu'}{}_{\mu\alpha} K^{\nu'\alpha}{}_{\nu}\Big)\Big(K^{\rho'\beta}{}_\sigma K^{\sigma'}{}_{\rho\beta}-K^{\rho'}{}_{\rho\beta} K^{\sigma'\beta}{}_{\sigma}\Big)\times\nonumber\\
&&\Big(K^{\tau'\gamma}{}_\eta K^{\eta'}{}_{\tau\gamma}-K^{\tau'}{}_{\tau\gamma} K^{\eta'\gamma}{}_{\eta}\Big) + \Big(\mathring{\nabla}_{\nu}K_{\sigma}{}^{\lambda}{}_{\mu} - 
  \mathring{\nabla}_{\sigma}K_{\nu}{}^{\lambda}{}_{\mu} +
  K_{\sigma}{}^{\rho}{}_{\mu}K_{\nu}{}^{\lambda}{}_{\rho} -
  K_{\sigma}{}^{\lambda}{}_{\rho}K_{\nu}{}^{\rho}{}_{\mu} \Big)\times\nonumber\\
  &&\Big(\mathring{\nabla}_{\beta}K_{\alpha}{}^{\sigma\nu} - 
  \mathring{\nabla}_{\alpha}K_{\beta}{}^{\sigma\nu} +
  K_{\alpha}{}^{\gamma\nu}K_{\beta}{}^{\sigma}{}_{\gamma} -
  K_{\alpha}{}^{\sigma}{}_{\gamma}K_{\beta}{}^{\gamma\nu}\Big)\Big(\mathring{\nabla}^{\mu}K_{\lambda}{}^{\alpha\beta} - 
  \mathring{\nabla}_{\lambda}K^{\mu\alpha\beta} +
  K_{\lambda}{}^{\eta\beta}K^{\mu\alpha}{}_{\eta} -
  K_{\lambda}{}^{\alpha}{}_{\eta}K^{\mu\eta\beta}\Big)\Big]\,,\nonumber\\
  \label{second}
\end{eqnarray}
which is the equivalent Teleparallel Lagrangian of the second string correction action~\eqref{second2}. For the term which is contracted with the Levi-Civita symbol, we have used that the contortion is antisymmetric in its last two indices, hence, no covariant derivative contributions of the contortion tensor appears there. Further higher other corrections such as $\mathcal{L}_{c}^{(3)},\mathcal{L}_{c}^{(4)},..$ will then also appear in a Teleparallel string effective action, and it might be expected that these corrections will contain the standard $\mathring{\mathcal{L}}_{c}^{(3)},\mathring{\mathcal{L}}_{c}^{(4)},..$ higher corrections based on the standard effective string theory. As a future work, it would be interesting to find out the correct and complete Teleparallel effective string action that would contain both the Gauss-Bonnet contribution studied in this paper, the second order contribution written above and further more terms that do not appear in the standard curvature based string effective action.\\

It would be interesting to consider the action~\eqref{action} without a scalar field and only with linear terms on all geometrical invariants. In fact this is analog of $L=\mathring{R}/(2\kappa^2)+\mathring{G}$ term but in our case we have several torsion analogs of the Gauss-Bonnet term. Generally speaking for such theory only $\mathring{R}$ (or its torsion equivalent $T$) gives the contribution to the field equations due to the fact that higher-order terms are topological invariants. However, if these terms are multiplied by coupling constants $c_i$ and these coupling constants are re-scaled as $c_i/(D-4)$ in the limit from $D \rightarrow 4$ of such torsional theory with linear topological invariants, we expect to find qualitatively novel rich black hole dynamics as it was just shown for $\mathring{R}/(2\kappa^2)+\alpha \mathring{G}/(D-4) $ theory in~\cite{Glavan:2019inb}. This is due to the fact that in the $D\rightarrow 4$ limit with singular rescaling of coupling constants of topological invariants, they give non-trivial contributions to the field equations. This will be considered elsewhere.\\

\par

\section*{Acknowledgements}
This article is based upon work from CANTATA COST (European Cooperation in Science and Technology) action CA15117, EU Framework Programme Horizon 2020. SB is supported by Mobilitas Pluss N$^\circ$ MOBJD423 by the Estonian government. PR acknowledges the Inter University Centre for Astronomy
and Astrophysics (IUCAA), Pune, India for granting visiting
associateship. MM would like to thank V.~Baran for support. SDO is partially supported by Russian Ministry of Science and High Education, project No FEWF-2020-003. The authors thank Tomi S. Koivisto for spotting an important typo in the manuscript.

\bibliographystyle{utphys}
\bibliography{references}

\providecommand{\href}[2]{#2}\begingroup\raggedright\begin{thebibliography}{10}

\bibitem{Capozziello:2011et}
S.~Capozziello and M.~De~Laurentis, ``{Extended Theories of Gravity},''
  \href{http://dx.doi.org/10.1016/j.physrep.2011.09.003}{{\em Phys. Rept.} {\bf
  509} (2011)  167--321},
\href{http://arxiv.org/abs/1108.6266}{{\tt arXiv:1108.6266 [gr-qc]}}.

\bibitem{Nojiri:2017ncd}
S.~Nojiri, S.~D. Odintsov, and V.~K. Oikonomou, ``{Modified Gravity Theories on
  a Nutshell: Inflation, Bounce and Late-time Evolution},''
  \href{http://dx.doi.org/10.1016/j.physrep.2017.06.001}{{\em Phys. Rept.} {\bf
  692} (2017)  1--104},
\href{http://arxiv.org/abs/1705.11098}{{\tt arXiv:1705.11098 [gr-qc]}}.

\bibitem{Nojiri:2010wj}
S.~Nojiri and S.~D. Odintsov, ``{Unified cosmic history in modified gravity:
  from F(R) theory to Lorentz non-invariant models},''
  \href{http://dx.doi.org/10.1016/j.physrep.2011.04.001}{{\em Phys. Rept.} {\bf
  505} (2011)  59--144},
\href{http://arxiv.org/abs/1011.0544}{{\tt arXiv:1011.0544 [gr-qc]}}.

\bibitem{Fradkin:1985ys}
E.~Fradkin and A.~A. Tseytlin, ``{Quantum String Theory Effective Action},''
  \href{http://dx.doi.org/10.1016/0550-3213(85)90559-0}{{\em Nucl. Phys. B}
  {\bf 261} (1985)  1--27}. [Erratum: Nucl.Phys.B 269, 745--745 (1986)].

\bibitem{Kawai:1998ab}
S.~Kawai, M.-a. Sakagami, and J.~Soda, ``{Instability of one loop superstring
  cosmology},'' \href{http://dx.doi.org/10.1016/S0370-2693(98)00925-3}{{\em
  Phys. Lett.} {\bf B437} (1998)  284--290},
\href{http://arxiv.org/abs/gr-qc/9802033}{{\tt arXiv:gr-qc/9802033 [gr-qc]}}.

\bibitem{Antoniadis:1993jc}
I.~Antoniadis, J.~Rizos, and K.~Tamvakis, ``{Singularity - free cosmological
  solutions of the superstring effective action},''
  \href{http://dx.doi.org/10.1016/0550-3213(94)90120-1}{{\em Nucl. Phys.} {\bf
  B415} (1994)  497--514},
\href{http://arxiv.org/abs/hep-th/9305025}{{\tt arXiv:hep-th/9305025
  [hep-th]}}.

\bibitem{Tseytlin:1991xk}
A.~A. Tseytlin and C.~Vafa, ``{Elements of string cosmology},''
  \href{http://dx.doi.org/10.1016/0550-3213(92)90327-8}{{\em Nucl. Phys.} {\bf
  B372} (1992)  443--466},
\href{http://arxiv.org/abs/hep-th/9109048}{{\tt arXiv:hep-th/9109048
  [hep-th]}}.

\bibitem{Brustein:1994kw}
R.~Brustein and G.~Veneziano, ``{The Graceful exit problem in string
  cosmology},'' \href{http://dx.doi.org/10.1016/0370-2693(94)91086-3}{{\em
  Phys. Lett.} {\bf B329} (1994)  429--434},
\href{http://arxiv.org/abs/hep-th/9403060}{{\tt arXiv:hep-th/9403060
  [hep-th]}}.

\bibitem{Easther:1995ba}
R.~Easther, K.-i. Maeda, and D.~Wands, ``{Tree level string cosmology},''
  \href{http://dx.doi.org/10.1103/PhysRevD.53.4247}{{\em Phys. Rev.} {\bf D53}
  (1996)  4247--4256},
\href{http://arxiv.org/abs/hep-th/9509074}{{\tt arXiv:hep-th/9509074
  [hep-th]}}.

\bibitem{Easther:1996yd}
R.~Easther and K.-i. Maeda, ``{One loop superstring cosmology and the
  nonsingular universe},''
  \href{http://dx.doi.org/10.1103/PhysRevD.54.7252}{{\em Phys. Rev.} {\bf D54}
  (1996)  7252--7260},
\href{http://arxiv.org/abs/hep-th/9605173}{{\tt arXiv:hep-th/9605173
  [hep-th]}}.

\bibitem{Antoniadis:1988vi}
I.~Antoniadis, C.~Bachas, J.~R. Ellis, and D.~V. Nanopoulos, ``{An Expanding
  Universe in String Theory},''
\href{http://dx.doi.org/10.1016/0550-3213(89)90095-3}{{\em Nucl. Phys.} {\bf
  B328} (1989)  117--139}.

\bibitem{Veneziano:1991ek}
G.~Veneziano, ``{Scale factor duality for classical and quantum strings},''
\href{http://dx.doi.org/10.1016/0370-2693(91)90055-U}{{\em Phys. Lett.} {\bf
  B265} (1991)  287--294}.

\bibitem{Gasperini:1994xg}
M.~Gasperini and G.~Veneziano, ``{Dilaton production in string cosmology},''
  \href{http://dx.doi.org/10.1103/PhysRevD.50.2519}{{\em Phys. Rev.} {\bf D50}
  (1994)  2519--2540},
\href{http://arxiv.org/abs/gr-qc/9403031}{{\tt arXiv:gr-qc/9403031 [gr-qc]}}.

\bibitem{Hwang:2005hb}
J.-c. Hwang and H.~Noh, ``{Classical evolution and quantum generation in
  generalized gravity theories including string corrections and tachyon:
  Unified analyses},'' \href{http://dx.doi.org/10.1103/PhysRevD.71.063536}{{\em
  Phys. Rev.} {\bf D71} (2005)  063536},
\href{http://arxiv.org/abs/gr-qc/0412126}{{\tt arXiv:gr-qc/0412126 [gr-qc]}}.

\bibitem{Elizalde:2007pi}
E.~Elizalde, S.~Jhingan, S.~Nojiri, S.~D. Odintsov, M.~Sami, and I.~Thongkool,
  ``{Dark energy generated from a (super)string effective action with higher
  order curvature corrections and a dynamical dilaton},''
  \href{http://dx.doi.org/10.1140/epjc/s10052-007-0463-8}{{\em Eur. Phys. J.}
  {\bf C53} (2008)  447--457},
\href{http://arxiv.org/abs/0705.1211}{{\tt arXiv:0705.1211 [hep-th]}}.

\bibitem{Kanti:1998jd}
P.~Kanti, J.~Rizos, and K.~Tamvakis, ``{Singularity free cosmological solutions
  in quadratic gravity},''
  \href{http://dx.doi.org/10.1103/PhysRevD.59.083512}{{\em Phys. Rev.} {\bf
  D59} (1999)  083512},
\href{http://arxiv.org/abs/gr-qc/9806085}{{\tt arXiv:gr-qc/9806085 [gr-qc]}}.

\bibitem{Maeda:2011zn}
K.-i. Maeda, N.~Ohta, and R.~Wakebe, ``{Accelerating Universes in String Theory
  via Field Redefinition},''
  \href{http://dx.doi.org/10.1140/epjc/s10052-012-1949-6}{{\em Eur. Phys. J.}
  {\bf C72} (2012)  1949},
\href{http://arxiv.org/abs/1111.3251}{{\tt arXiv:1111.3251 [hep-th]}}.

\bibitem{Guo:2009uk}
Z.-K. Guo and D.~J. Schwarz, ``{Power spectra from an inflaton coupled to the
  Gauss-Bonnet term},''
  \href{http://dx.doi.org/10.1103/PhysRevD.80.063523}{{\em Phys. Rev.} {\bf
  D80} (2009)  063523},
\href{http://arxiv.org/abs/0907.0427}{{\tt arXiv:0907.0427 [hep-th]}}.

\bibitem{Guo:2010jr}
Z.-K. Guo and D.~J. Schwarz, ``{Slow-roll inflation with a Gauss-Bonnet
  correction},'' \href{http://dx.doi.org/10.1103/PhysRevD.81.123520}{{\em Phys.
  Rev.} {\bf D81} (2010)  123520},
\href{http://arxiv.org/abs/1001.1897}{{\tt arXiv:1001.1897 [hep-th]}}.

\bibitem{Jiang:2013gza}
P.-X. Jiang, J.-W. Hu, and Z.-K. Guo, ``{Inflation coupled to a Gauss-Bonnet
  term},'' \href{http://dx.doi.org/10.1103/PhysRevD.88.123508}{{\em Phys. Rev.}
  {\bf D88} (2013)  123508},
\href{http://arxiv.org/abs/1310.5579}{{\tt arXiv:1310.5579 [hep-th]}}.

\bibitem{Kanti:2015pda}
P.~Kanti, R.~Gannouji, and N.~Dadhich, ``{Gauss-Bonnet Inflation},''
  \href{http://dx.doi.org/10.1103/PhysRevD.92.041302}{{\em Phys. Rev.} {\bf
  D92} (2015) no.~4, 041302},
\href{http://arxiv.org/abs/1503.01579}{{\tt arXiv:1503.01579 [hep-th]}}.

\bibitem{Nozari:2017rta}
K.~Nozari and N.~Rashidi, ``{Perturbation, non-Gaussianity, and reheating in a
  Gauss-Bonnet $\alpha$-attractor model},''
  \href{http://dx.doi.org/10.1103/PhysRevD.95.123518}{{\em Phys. Rev.} {\bf
  D95} (2017) no.~12, 123518},
\href{http://arxiv.org/abs/1705.02617}{{\tt arXiv:1705.02617 [astro-ph.CO]}}.

\bibitem{Chakraborty:2018scm}
S.~Chakraborty, T.~Paul, and S.~SenGupta, ``{Inflation driven by
  Einstein-Gauss-Bonnet gravity},''
  \href{http://dx.doi.org/10.1103/PhysRevD.98.083539}{{\em Phys. Rev.} {\bf
  D98} (2018) no.~8, 083539},
\href{http://arxiv.org/abs/1804.03004}{{\tt arXiv:1804.03004 [gr-qc]}}.

\bibitem{Odintsov:2018zhw}
S.~D. Odintsov and V.~K. Oikonomou, ``{Viable Inflation in Scalar-Gauss-Bonnet
  Gravity and Reconstruction from Observational Indices},''
  \href{http://dx.doi.org/10.1103/PhysRevD.98.044039}{{\em Phys. Rev.} {\bf
  D98} (2018) no.~4, 044039},
\href{http://arxiv.org/abs/1808.05045}{{\tt arXiv:1808.05045 [gr-qc]}}.

\bibitem{Yi:2018dhl}
Z.~Yi and Y.~Gong, ``{Gauss–Bonnet Inflation and the String Swampland},''
  \href{http://dx.doi.org/10.3390/universe5090200}{{\em Universe} {\bf 5}
  (2019) no.~9, 200},
\href{http://arxiv.org/abs/1811.01625}{{\tt arXiv:1811.01625 [gr-qc]}}.

\bibitem{vandeBruck:2016xvt}
C.~van~de Bruck, K.~Dimopoulos, and C.~Longden, ``{Reheating in
  Gauss-Bonnet-coupled inflation},''
  \href{http://dx.doi.org/10.1103/PhysRevD.94.023506}{{\em Phys. Rev.} {\bf
  D94} (2016) no.~2, 023506},
\href{http://arxiv.org/abs/1605.06350}{{\tt arXiv:1605.06350 [astro-ph.CO]}}.

\bibitem{Cai:2015emx}
Y.-F. Cai, S.~Capozziello, M.~De~Laurentis, and E.~N. Saridakis, ``{f(T)
  teleparallel gravity and cosmology},''
  \href{http://dx.doi.org/10.1088/0034-4885/79/10/106901}{{\em Rept. Prog.
  Phys.} {\bf 79} (2016) no.~10, 106901},
\href{http://arxiv.org/abs/1511.07586}{{\tt arXiv:1511.07586 [gr-qc]}}.

\bibitem{Ferraro:2006jd}
R.~Ferraro and F.~Fiorini, ``{Modified teleparallel gravity: Inflation without
  inflaton},'' \href{http://dx.doi.org/10.1103/PhysRevD.75.084031}{{\em Phys.
  Rev.} {\bf D75} (2007)  084031},
\href{http://arxiv.org/abs/gr-qc/0610067}{{\tt arXiv:gr-qc/0610067 [gr-qc]}}.

\bibitem{Ferraro:2008ey}
R.~Ferraro and F.~Fiorini, ``{On Born-Infeld Gravity in Weitzenbock
  spacetime},'' \href{http://dx.doi.org/10.1103/PhysRevD.78.124019}{{\em Phys.
  Rev.} {\bf D78} (2008)  124019},
\href{http://arxiv.org/abs/0812.1981}{{\tt arXiv:0812.1981 [gr-qc]}}.

\bibitem{Bamba:2010wb}
K.~Bamba, C.-Q. Geng, C.-C. Lee, and L.-W. Luo, ``{Equation of state for dark
  energy in $f(T)$ gravity},''
  \href{http://dx.doi.org/10.1088/1475-7516/2011/01/021}{{\em JCAP} {\bf 1101}
  (2011)  021},
\href{http://arxiv.org/abs/1011.0508}{{\tt arXiv:1011.0508 [astro-ph.CO]}}.

\bibitem{Dent:2011zz}
J.~B. Dent, S.~Dutta, and E.~N. Saridakis, ``{f(T) gravity mimicking dynamical
  dark energy. Background and perturbation analysis},''
  \href{http://dx.doi.org/10.1088/1475-7516/2011/01/009}{{\em JCAP} {\bf 1101}
  (2011)  009},
\href{http://arxiv.org/abs/1010.2215}{{\tt arXiv:1010.2215 [astro-ph.CO]}}.

\bibitem{Wu:2010av}
P.~Wu and H.~W. Yu, ``{$f(T)$ models with phantom divide line crossing},''
  \href{http://dx.doi.org/10.1140/epjc/s10052-011-1552-2}{{\em Eur. Phys. J.}
  {\bf C71} (2011)  1552},
\href{http://arxiv.org/abs/1008.3669}{{\tt arXiv:1008.3669 [gr-qc]}}.

\bibitem{Hohmann:2017jao}
M.~Hohmann, L.~Jarv, and U.~Ualikhanova, ``{Dynamical systems approach and
  generic properties of $f(T)$ cosmology},''
  \href{http://dx.doi.org/10.1103/PhysRevD.96.043508}{{\em Phys. Rev.} {\bf
  D96} (2017) no.~4, 043508},
\href{http://arxiv.org/abs/1706.02376}{{\tt arXiv:1706.02376 [gr-qc]}}.

\bibitem{Nunes:2018xbm}
R.~C. Nunes, ``{Structure formation in $f(T)$ gravity and a solution for $H_0$
  tension},'' \href{http://dx.doi.org/10.1088/1475-7516/2018/05/052}{{\em JCAP}
  {\bf 1805} (2018) no.~05, 052},
\href{http://arxiv.org/abs/1802.02281}{{\tt arXiv:1802.02281 [gr-qc]}}.

\bibitem{Cai:2011tc}
Y.-F. Cai, S.-H. Chen, J.~B. Dent, S.~Dutta, and E.~N. Saridakis, ``{Matter
  Bounce Cosmology with the f(T) Gravity},''
  \href{http://dx.doi.org/10.1088/0264-9381/28/21/215011}{{\em Class. Quant.
  Grav.} {\bf 28} (2011)  215011},
\href{http://arxiv.org/abs/1104.4349}{{\tt arXiv:1104.4349 [astro-ph.CO]}}.

\bibitem{Cai:2018rzd}
Y.-F. Cai, C.~Li, E.~N. Saridakis, and L.~Xue, ``{$f(T)$ gravity after GW170817
  and GRB170817A},'' \href{http://dx.doi.org/10.1103/PhysRevD.97.103513}{{\em
  Phys. Rev.} {\bf D97} (2018) no.~10, 103513},
\href{http://arxiv.org/abs/1801.05827}{{\tt arXiv:1801.05827 [gr-qc]}}.

\bibitem{Capozziello:2019cav}
S.~Capozziello, R.~D'Agostino, and O.~Luongo, ``{Extended Gravity
  Cosmography},'' \href{http://dx.doi.org/10.1142/S0218271819300167}{{\em Int.
  J. Mod. Phys.} {\bf D28} (2019) no.~10, 1930016},
\href{http://arxiv.org/abs/1904.01427}{{\tt arXiv:1904.01427 [gr-qc]}}.

\bibitem{Kofinas:2014owa}
G.~Kofinas and E.~N. Saridakis, ``{Teleparallel equivalent of Gauss-Bonnet
  gravity and its modifications},''
  \href{http://dx.doi.org/10.1103/PhysRevD.90.084044}{{\em Phys. Rev.} {\bf
  D90} (2014)  084044},
\href{http://arxiv.org/abs/1404.2249}{{\tt arXiv:1404.2249 [gr-qc]}}.

\bibitem{Bahamonde:2015zma}
S.~Bahamonde, C.~G. Böhmer, and M.~Wright, ``{Modified teleparallel theories
  of gravity},'' \href{http://dx.doi.org/10.1103/PhysRevD.92.104042}{{\em Phys.
  Rev.} {\bf D92} (2015) no.~10, 104042},
\href{http://arxiv.org/abs/1508.05120}{{\tt arXiv:1508.05120 [gr-qc]}}.

\bibitem{Bahamonde:2017wwk}
S.~Bahamonde, C.~G. Böhmer, and M.~Krššák, ``{New classes of modified
  teleparallel gravity models},''
  \href{http://dx.doi.org/10.1016/j.physletb.2017.10.026}{{\em Phys. Lett.}
  {\bf B775} (2017)  37--43},
\href{http://arxiv.org/abs/1706.04920}{{\tt arXiv:1706.04920 [gr-qc]}}.

\bibitem{Kofinas:2014daa}
G.~Kofinas and E.~N. Saridakis, ``{Cosmological applications of $F(T,T_G)$
  gravity},'' \href{http://dx.doi.org/10.1103/PhysRevD.90.084045}{{\em Phys.
  Rev.} {\bf D90} (2014)  084045},
\href{http://arxiv.org/abs/1408.0107}{{\tt arXiv:1408.0107 [gr-qc]}}.

\bibitem{Kofinas:2014aka}
G.~Kofinas, G.~Leon, and E.~N. Saridakis, ``{Dynamical behavior in $f(T,T_G)$
  cosmology},'' \href{http://dx.doi.org/10.1088/0264-9381/31/17/175011}{{\em
  Class. Quant. Grav.} {\bf 31} (2014)  175011},
\href{http://arxiv.org/abs/1404.7100}{{\tt arXiv:1404.7100 [gr-qc]}}.

\bibitem{Bahamonde:2016kba}
S.~Bahamonde and C.~G. Böhmer, ``{Modified teleparallel theories of gravity:
  Gauss–Bonnet and trace extensions},''
  \href{http://dx.doi.org/10.1140/epjc/s10052-016-4419-8}{{\em Eur. Phys. J.}
  {\bf C76} (2016) no.~10, 578},
\href{http://arxiv.org/abs/1606.05557}{{\tt arXiv:1606.05557 [gr-qc]}}.

\bibitem{Gonzalez:2019tky}
P.~A. González, S.~Reyes, and Y.~Vásquez, ``{Teleparallel Equivalent of
  Lovelock Gravity, Generalizations and Cosmological Applications},''
  \href{http://dx.doi.org/10.1088/1475-7516/2019/07/040}{{\em JCAP} {\bf 1907}
  (2019)  040},
\href{http://arxiv.org/abs/1905.07633}{{\tt arXiv:1905.07633 [gr-qc]}}.

\bibitem{Bahamonde:2015hza}
S.~Bahamonde and M.~Wright, ``{Teleparallel quintessence with a nonminimal
  coupling to a boundary term},''
  \href{http://dx.doi.org/10.1103/PhysRevD.92.084034,
  10.1103/PhysRevD.93.109901}{{\em Phys. Rev.} {\bf D92} (2015) no.~8, 084034},
  \href{http://arxiv.org/abs/1508.06580}{{\tt arXiv:1508.06580 [gr-qc]}}.
[Erratum: Phys. Rev.D93,no.10,109901(2016)].

\bibitem{Zubair:2016uhx}
M.~Zubair, S.~Bahamonde, and M.~Jamil, ``{Generalized Second Law of
  Thermodynamic in Modified Teleparallel Theory},''
  \href{http://dx.doi.org/10.1140/epjc/s10052-017-5043-y}{{\em Eur. Phys. J.}
  {\bf C77} (2017) no.~7, 472},
\href{http://arxiv.org/abs/1604.02996}{{\tt arXiv:1604.02996 [gr-qc]}}.

\bibitem{Uzan:1999ch}
J.-P. Uzan, ``{Cosmological scaling solutions of nonminimally coupled scalar
  fields},'' \href{http://dx.doi.org/10.1103/PhysRevD.59.123510}{{\em Phys.
  Rev.} {\bf D59} (1999)  123510},
\href{http://arxiv.org/abs/gr-qc/9903004}{{\tt arXiv:gr-qc/9903004 [gr-qc]}}.

\bibitem{Bartolo:1999sq}
N.~Bartolo and M.~Pietroni, ``{Scalar tensor gravity and quintessence},''
  \href{http://dx.doi.org/10.1103/PhysRevD.61.023518}{{\em Phys. Rev.} {\bf
  D61} (2000)  023518},
\href{http://arxiv.org/abs/hep-ph/9908521}{{\tt arXiv:hep-ph/9908521
  [hep-ph]}}.

\bibitem{Bahamonde:2018miw}
S.~Bahamonde, M.~Marciu, and P.~Rudra, ``{Generalised teleparallel quintom dark
  energy non-minimally coupled with the scalar torsion and a boundary term},''
  \href{http://dx.doi.org/10.1088/1475-7516/2018/04/056}{{\em JCAP} {\bf 1804}
  (2018) no.~04, 056},
\href{http://arxiv.org/abs/1802.09155}{{\tt arXiv:1802.09155 [gr-qc]}}.

\bibitem{Bahamonde:2017bps}
S.~Bahamonde, S.~Capozziello, M.~Faizal, and R.~C. Nunes, ``{Nonlocal
  Teleparallel Cosmology},''
  \href{http://dx.doi.org/10.1140/epjc/s10052-017-5210-1}{{\em Eur. Phys. J.}
  {\bf C77} (2017) no.~9, 628},
\href{http://arxiv.org/abs/1709.02692}{{\tt arXiv:1709.02692 [gr-qc]}}.

\bibitem{Bahamonde:2017sdo}
S.~Bahamonde, S.~Capozziello, and K.~F. Dialektopoulos, ``{Constraining
  Generalized Non-local Cosmology from Noether Symmetries},''
  \href{http://dx.doi.org/10.1140/epjc/s10052-017-5283-x}{{\em Eur. Phys. J.}
  {\bf C77} (2017) no.~11, 722},
\href{http://arxiv.org/abs/1708.06310}{{\tt arXiv:1708.06310 [gr-qc]}}.

\bibitem{Bahamonde:2019gjk}
S.~Bahamonde, M.~Marciu, and J.~L. Said, ``{Generalized Tachyonic Teleparallel
  cosmology},'' \href{http://dx.doi.org/10.1140/epjc/s10052-019-6833-1}{{\em
  Eur. Phys. J.} {\bf C79} (2019) no.~4, 324},
\href{http://arxiv.org/abs/1901.04973}{{\tt arXiv:1901.04973 [gr-qc]}}.

\bibitem{Marciu:2017sji}
M.~Marciu, ``{Dynamical properties of scaling solutions in teleparallel dark
  energy cosmologies with nonminimal coupling},''
\href{http://dx.doi.org/10.1142/S0218271817501036}{{\em Int. J. Mod. Phys.}
  {\bf D26} (2017) no.~09, 1750103}.

\bibitem{Gecim:2017hmn}
G.~Gecim and Y.~Kucukakca, ``{Scalar–tensor teleparallel gravity with
  boundary term by Noether symmetries},''
  \href{http://dx.doi.org/10.1142/S0219887818501517}{{\em Int. J. Geom. Meth.
  Mod. Phys.} {\bf 15} (2018) no.~09, 1850151},
\href{http://arxiv.org/abs/1708.07430}{{\tt arXiv:1708.07430 [gr-qc]}}.

\bibitem{Bahamonde:2019shr}
S.~Bahamonde, K.~F. Dialektopoulos, and J.~Levi~Said, ``{Can Horndeski Theory
  be recast using Teleparallel Gravity?},''
  \href{http://dx.doi.org/10.1103/PhysRevD.100.064018}{{\em Phys. Rev.} {\bf
  D100} (2019) no.~6, 064018},
\href{http://arxiv.org/abs/1904.10791}{{\tt arXiv:1904.10791 [gr-qc]}}.

\bibitem{Bahamonde:2019ipm}
S.~Bahamonde, K.~F. Dialektopoulos, V.~Gakis, and J.~Levi~Said, ``{Reviving
  Horndeski Theory using Teleparallel Gravity after GW170817},''
\href{http://arxiv.org/abs/1907.10057}{{\tt arXiv:1907.10057 [gr-qc]}}.

\bibitem{Krssak:2015oua}
M.~Krššák and E.~N. Saridakis, ``{The covariant formulation of f(T)
  gravity},'' \href{http://dx.doi.org/10.1088/0264-9381/33/11/115009}{{\em
  Class. Quant. Grav.} {\bf 33} (2016) no.~11, 115009},
\href{http://arxiv.org/abs/1510.08432}{{\tt arXiv:1510.08432 [gr-qc]}}.

\bibitem{Krssak:2018ywd}
M.~Krssak, R.~J. van~den Hoogen, J.~G. Pereira, C.~G. Böhmer, and A.~A. Coley,
  ``{Teleparallel theories of gravity: illuminating a fully invariant
  approach},'' \href{http://dx.doi.org/10.1088/1361-6382/ab2e1f}{{\em Class.
  Quant. Grav.} {\bf 36} (2019) no.~18, 183001},
\href{http://arxiv.org/abs/1810.12932}{{\tt arXiv:1810.12932 [gr-qc]}}.

\bibitem{Tamanini:2012hg}
N.~Tamanini and C.~G. Boehmer, ``{Good and bad tetrads in f(T) gravity},''
  \href{http://dx.doi.org/10.1103/PhysRevD.86.044009}{{\em Phys. Rev.} {\bf
  D86} (2012)  044009},
\href{http://arxiv.org/abs/1204.4593}{{\tt arXiv:1204.4593 [gr-qc]}}.

\bibitem{Beltran:2015hja}
S.~Bahamonde, C.~G. Böhmer, F.~S.~N. Lobo, and D.~Sáez-Gómez, ``{Generalized
  $f(R,\phi,X)$ Gravity and the Late-Time Cosmic Acceleration},''
  \href{http://dx.doi.org/10.3390/universe1020186}{{\em Universe} {\bf 1}
  (2015) no.~2, 186--198},
\href{http://arxiv.org/abs/1506.07728}{{\tt arXiv:1506.07728 [gr-qc]}}.

\bibitem{Nojiri:2005vv}
S.~Nojiri, S.~D. Odintsov, and M.~Sasaki, ``{Gauss-Bonnet dark energy},''
  \href{http://dx.doi.org/10.1103/PhysRevD.71.123509}{{\em Phys. Rev.} {\bf
  D71} (2005)  123509},
\href{http://arxiv.org/abs/hep-th/0504052}{{\tt arXiv:hep-th/0504052
  [hep-th]}}.

\bibitem{Bahamonde:2018ibz}
S.~Bahamonde, U.~Camci, and S.~Capozziello, ``{Noether symmetries and boundary
  terms in extended Teleparallel gravity cosmology},''
  \href{http://dx.doi.org/10.1088/1361-6382/ab0510}{{\em Class. Quant. Grav.}
  {\bf 36} (2019) no.~6, 065013},
\href{http://arxiv.org/abs/1807.02891}{{\tt arXiv:1807.02891 [gr-qc]}}.

\bibitem{Cognola:2006eg}
G.~Cognola, E.~Elizalde, S.~Nojiri, S.~D. Odintsov, and S.~Zerbini, ``{Dark
  energy in modified Gauss-Bonnet gravity: Late-time acceleration and the
  hierarchy problem},''
  \href{http://dx.doi.org/10.1103/PhysRevD.73.084007}{{\em Phys. Rev.} {\bf
  D73} (2006)  084007},
\href{http://arxiv.org/abs/hep-th/0601008}{{\tt arXiv:hep-th/0601008
  [hep-th]}}.

\bibitem{Geng:2011aj}
C.-Q. Geng, C.-C. Lee, E.~N. Saridakis, and Y.-P. Wu, ``{“Teleparallel”
  dark energy},'' \href{http://dx.doi.org/10.1016/j.physletb.2011.09.082}{{\em
  Phys. Lett.} {\bf B704} (2011)  384--387},
\href{http://arxiv.org/abs/1109.1092}{{\tt arXiv:1109.1092 [hep-th]}}.

\bibitem{Xu:2012jf}
C.~Xu, E.~N. Saridakis, and G.~Leon, ``{Phase-Space analysis of Teleparallel
  Dark Energy},'' \href{http://dx.doi.org/10.1088/1475-7516/2012/07/005}{{\em
  JCAP} {\bf 1207} (2012)  005},
\href{http://arxiv.org/abs/1202.3781}{{\tt arXiv:1202.3781 [gr-qc]}}.

\bibitem{Gross:1986mw}
D.~J. Gross and J.~H. Sloan, ``{The Quartic Effective Action for the Heterotic
  String},'' \href{http://dx.doi.org/10.1016/0550-3213(87)90465-2}{{\em Nucl.
  Phys. B} {\bf 291} (1987)  41--89}.

\bibitem{Metsaev:1987ju}
R.~Metsaev and A.~A. Tseytlin, ``{ON LOOP CORRECTIONS TO STRING THEORY
  EFFECTIVE ACTIONS},''
  \href{http://dx.doi.org/10.1016/0550-3213(88)90306-9}{{\em Nucl. Phys. B}
  {\bf 298} (1988)  109--132}.

\bibitem{Mavromatos:2000az}
N.~E. Mavromatos and J.~Rizos, ``{String inspired higher curvature terms and
  the Randall-Sundrum scenario},''
  \href{http://dx.doi.org/10.1103/PhysRevD.62.124004}{{\em Phys. Rev. D} {\bf
  62} (2000)  124004}, \href{http://arxiv.org/abs/hep-th/0008074}{{\tt
  arXiv:hep-th/0008074}}.

\bibitem{Bahamonde:2017ize}
S.~Bahamonde, C.~G. Böhmer, S.~Carloni, E.~J. Copeland, W.~Fang, and
  N.~Tamanini, ``{Dynamical systems applied to cosmology: dark energy and
  modified gravity},''
  \href{http://dx.doi.org/10.1016/j.physrep.2018.09.001}{{\em Phys. Rept.} {\bf
  775-777} (2018)  1--122},
\href{http://arxiv.org/abs/1712.03107}{{\tt arXiv:1712.03107 [gr-qc]}}.

\bibitem{Alam:2003sc}
U.~Alam, V.~Sahni, T.~D. Saini, and A.~A. Starobinsky, ``{Exploring the
  expanding universe and dark energy using the Statefinder diagnostic},''
  \href{http://dx.doi.org/10.1046/j.1365-8711.2003.06871.x}{{\em Mon. Not. Roy.
  Astron. Soc.} {\bf 344} (2003)  1057},
\href{http://arxiv.org/abs/astro-ph/0303009}{{\tt arXiv:astro-ph/0303009
  [astro-ph]}}.

\bibitem{Sahni:2002fz}
V.~Sahni, T.~D. Saini, A.~A. Starobinsky, and U.~Alam, ``{Statefinder: A New
  geometrical diagnostic of dark energy},''
  \href{http://dx.doi.org/10.1134/1.1574831}{{\em JETP Lett.} {\bf 77} (2003)
  201--206}, \href{http://arxiv.org/abs/astro-ph/0201498}{{\tt
  arXiv:astro-ph/0201498 [astro-ph]}}.
[Pisma Zh. Eksp. Teor. Fiz.77,249(2003)].

\bibitem{Lykkas:2015kls}
A.~Lykkas and L.~Perivolaropoulos, ``{Scalar-Tensor Quintessence with a linear
  potential: Avoiding the Big Crunch cosmic doomsday},''
  \href{http://dx.doi.org/10.1103/PhysRevD.93.043513}{{\em Phys. Rev.} {\bf
  D93} (2016) no.~4, 043513},
\href{http://arxiv.org/abs/1511.08732}{{\tt arXiv:1511.08732 [gr-qc]}}.

\bibitem{Perivolaropoulos:2004yr}
L.~Perivolaropoulos, ``{Constraints on linear negative potentials in
  quintessence and phantom models from recent supernova data},''
  \href{http://dx.doi.org/10.1103/PhysRevD.71.063503}{{\em Phys. Rev.} {\bf
  D71} (2005)  063503},
\href{http://arxiv.org/abs/astro-ph/0412308}{{\tt arXiv:astro-ph/0412308
  [astro-ph]}}.

\bibitem{Bento:1988sr}
M.~C. Bento, O.~Bertolami, A.~B. Henriques, and J.~C. Romao, ``{Order
  $\alpha^{\prime 2}$ Terms in the Gravitational Sector of String Effective
  Actions With the Inclusion of the Dilaton Field},''
\href{http://dx.doi.org/10.1016/0370-2693(89)91412-3}{{\em Phys. Lett.} {\bf
  B218} (1989)  162--168}.

\bibitem{Glavan:2019inb}
D.~Glavan and C.~Lin, ``{Einstein-Gauss-Bonnet gravity in 4-dimensional
  space-time},'' \href{http://dx.doi.org/10.1103/PhysRevLett.124.081301}{{\em
  Phys. Rev. Lett.} {\bf 124} (2020) no.~8, 081301},
\href{http://arxiv.org/abs/1905.03601}{{\tt arXiv:1905.03601 [gr-qc]}}.

\end{thebibliography}\endgroup

\end{document}